\documentclass[sigconf,natbib=false]{acmart}
\AtBeginDocument{%
  }

\setcopyright{acmlicensed}
\copyrightyear{2018}
\acmYear{2018}
\acmDOI{XXXXXXX.XXXXXXX}
\acmConference[Conference acronym 'XX]{Make sure to enter the correct
  conference title from your rights confirmation email}{June 03--05,
  2018}{Woodstock, NY}
\acmISBN{978-1-4503-XXXX-X/2018/06}

\RequirePackage[
  datamodel=acmdatamodel,
  style=acmnumeric,
  ]{biblatex}

\graphicspath{{figs/}{figures/}{pictures/}{images/}{./}} 

\usepackage{tabu}                      
\usepackage{booktabs}                  
\usepackage{mwe}                       

\usepackage{amsmath,amsfonts}
\usepackage{algorithmic}
\usepackage{algorithm}
\usepackage{array}
\usepackage[caption=false,font=normalsize,labelfont=sf,textfont=sf]{subfig}
\usepackage{textcomp}
\usepackage{stfloats}
\usepackage{url}
\usepackage{verbatim}
\usepackage{graphicx}
\usepackage{float}
\usepackage{soul}
\usepackage{fontawesome5}
\usepackage[para,online,flushleft]{threeparttable}

\usepackage{pgf}
\usepackage{pgfplots}
\pgfplotsset{compat=1.18}
\usepackage{tikz}
\usepackage{tikzscale}
\usetikzlibrary{arrows,external}
\usetikzlibrary{spy}
\usepackage{pgfplots}
\usepackage{tikz-dimline}
\usetikzlibrary{positioning}
\usetikzlibrary{arrows.meta}
\usetikzlibrary{graphs, quotes}
\usetikzlibrary{decorations.pathreplacing, calligraphy}
\usetikzlibrary{matrix}
\usepgfplotslibrary{statistics}
\usetikzlibrary{patterns}

\pgfplotsset{ 
    /pgfplots/boxplot/average={auto},
    boxplot/every average/.style={/tikz/mark=star,}
    }
\usepackage{soul}

\begin{document}

\title{Holoview: An Immersive Mixed-Reality Visualization System for Anatomical Education}

\author{Anshul Goswami}
\affiliation{%
  \institution{Indraprastha Institute of Information Technology Delhi}
  \city{New Delhi}
  \country{India}}
\email{anshul20361@iiitd.ac.in}

\author{Ojaswa Sharma}
\affiliation{%
  \institution{Indraprastha Institute of Information Technology Delhi}
  \city{New Delhi}
  \country{India}}
\email{ojaswa@iiitd.ac.in}

\renewcommand{\shortauthors}{Goswami and Sharma}


\begin{abstract}
We present Holoview, an augmented reality (AR) system designed to support immersive and interactive learning of human anatomy. Holoview enables users to dynamically explore volumetric anatomical data through intuitive hand gestures in a 3D AR environment, allowing inspection of individual organs and cross-sectional views via clipping and bioscope features. The system adopts a lightweight client–server architecture optimized for real-time performance on the HoloLens through hybrid and foveated rendering. Our user study demonstrated Holoview’s educational effectiveness, with participants showing a 135\% improvement in task-specific knowledge and reporting increased confidence in understanding anatomical structures. The system was perceived as engaging and intuitive, particularly for organ selection and cross-sectional exploration, with low cognitive load and increasing ease of use over time. These findings highlight Holoview's potential to enhance anatomy learning through immersive, user-centered AR experiences.

\end{abstract}  


\begin{teaserfigure}
  \centering
  \includegraphics[width=\linewidth, alt={Holoview}]{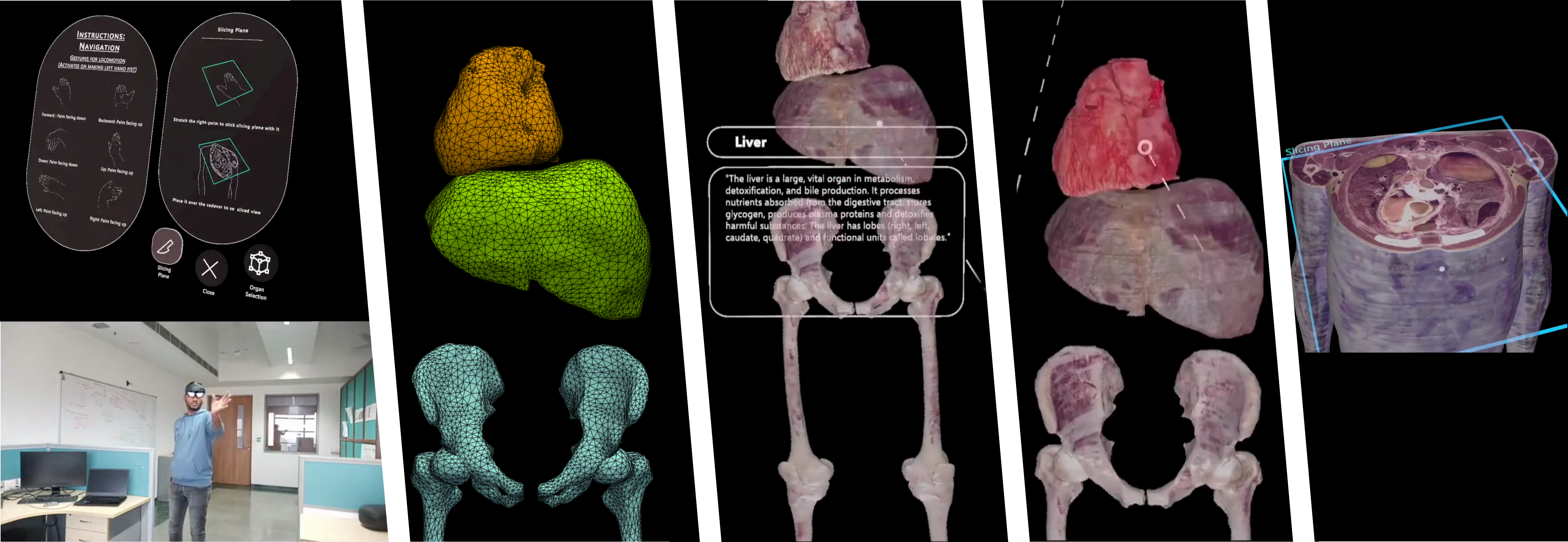}
  \caption{Holoview is an immersive augmented reality (AR) framework designed for interactive anatomy visualization and exploration. It features intuitive hand-gesture controls, real-time stereoscopic rendering, and dynamic slicing of volumetric medical data, enabling users to navigate and interact with anatomical structures seamlessly. By using remote visualization and hybrid mesh-based rendering, Holoview ensures high-performance visualization on lightweight AR devices like the HoloLens, making it a powerful tool for medical education.}
  \label{fig:teaser}
\end{teaserfigure}

\received{20 February 2007}
\received[revised]{12 March 2009}
\received[accepted]{5 June 2009}

\maketitle

\section{Introduction}
Augmented Reality (AR) technology has significantly transformed medical education by introducing innovative tools that transcend the limitations of traditional learning methods. AR headsets have emerged as powerful devices in the medical field, providing clinicians and medical students with enhanced insights into patient data and anatomical structures. This paper presents a novel AR application specifically designed to address the growing demand for advanced educational tools, aiming to revolutionize the study of human anatomy for students. 

In contemporary medical practice, AR applications are widely utilized for diverse purposes, such as anatomy training, patient monitoring, image-guided interventions, and surgical navigation. However, many existing AR systems exhibit notable limitations, particularly in their ability to represent human anatomy comprehensively. These systems \cite{MDcave,spatialTouch} often rely on auxiliary devices like desktops with mouse and keyboard setups or large touchscreens for visualization control. While these configurations address visualization challenges, they reduce immersion by requiring additional external devices, which complicates workflows and burdens the user.  Recent advances in AR technology have resulted in lighter devices, reducing neck strain and discomfort during extended use while enhancing immersion.

The AR framework \emph{Holoview} introduced in this work is designed to provide a fully immersive experience without requiring auxiliary tools. Recognizing the computational limitations of AR hardware, our system offloads intensive processing tasks to a remote server, ensuring high performance without compromising device usability. To further enhance the user experience, hand gestures have been incorporated as the primary mechanism, allowing users to interact seamlessly with the system. Traditional anatomy learning methods, such as textbooks and videos, offer a foundational understanding but confine learners to two-dimensional representations, limiting the depth of exploration. Our AR framework enables medical students to navigate the human body and interact with anatomical structures in an immersive and interactive environment, offering new opportunities for exploration and engagement. The significance of this work lies in its ability to transform anatomy education by offering an intuitive, interactive, and accessible learning tool. Key contributions of Holoview are:
\begin{itemize}
  \item An immersive AR framework for interactive 3D anatomy visualization, enabling dynamic selection, inspection, and navigation of organs within volumetric data through natural hand gestures,
  \item A high-performance remote rendering pipeline that leverages hybrid mesh-volume rendering, foveated encoding, and multi-threaded communication, delivering stereoscopic visualization on lightweight clients like the HoloLens, and
  \item A gesture-driven interaction design tailored for anatomy education, supporting intuitive modes such as selective rendering, slicing via a clipping plane, and detailed inspection using a bioscope tool.
\end{itemize}

\section{Related Work}

\subsection{Visualization in Medical Education}
Visualization plays a crucial role in medical education by enabling learners to understand complex anatomical and physiological structures through graphical representations. Traditional methods, such as textbooks and static models, have been supplemented by physical models, digital visualizations, including 3D models, virtual reality, and interactive simulations, which enhance spatial understanding and retention. Schindler et al. \cite{schindler2020anatomical} introduce a low-cost workflow for creating interactive 2D and 3D anatomical models using printed paper and colored filters. It aims to enhance anatomical education and engagement through tangible, color-based exploration of human structures. In the context of physicalization for medical education, Pahr et al. \cite{pahr2021vologram} introduced the Vologram approach, which utilizes a slide-based interactive sculpture to facilitate the exploration of human anatomy through tangible and spatially organized visual representations.

\subsection{Augmented Reality in Medical Training}
AR technologies have been identified as significant educational tools in medical training. Barsom et al.~\cite{Barsom2016} conducted a systematic review to demonstrate the effectiveness of AR applications in improving medical training outcomes, particularly in surgical procedures and anatomy education. The authors noted that Head-Mounted Displays (HMDs) enable repeated practice without ethical concerns associated with traditional methods, thus facilitating a more engaging learning environment. Barteit et al.~\cite{barteit2021} emphasized the advantages of AR and Mixed Reality (MR) beyond surgical applications, highlighting their effectiveness in teaching complex subjects like anatomy and anesthesia. Their findings suggest that AR can make intricate medical concepts more accessible, thereby enhancing student comprehension and retention.

Ma et al.~\cite{Ma2016} introduced a personalized AR system that enables users to visualize anatomical data overlayed on their own bodies, enhancing engagement and understanding. This personalized approach aligns with the findings of Tang et al.~\cite{Tang2020}, who reported that AR technologies improve educational outcomes by providing students with interactive experiences. Additionally, Moro et al.~\cite{Moro2021} found that three-dimensional visualization methods, including AR, are preferred by students and are more effective than traditional teaching methods in anatomy education. 

Additionally, Schott et al.~\cite{schoot_2021} have highlighted how Microsoft HoloLens AR applications can be used for stereoscopic visualization of 3D data, especially in volumetric medical imaging, providing an improved understanding of three-dimensional structures. Integrating AR with techniques such as cinematic volume rendering emerges as a promising synergy \cite{niedermayr2024}, which can revolutionize medical imaging by providing photo-realistic and interactive 3D visualizations of the imaging data. This integration allows for hyper-realistic depictions of deep human anatomy, improving results and elevating the medical training and learning experience.

\subsection{Interactive Exploration of Volumetric Data}
Augmented Reality navigation systems have been increasingly integrated into medical procedures, particularly in endoscopy to enhance exploration and improve patient outcomes. Navigation has been one of the most important innovations in the area of medical examination since the invention of endoscopy. Endoscopy has improved the visualization and diagnosis of the interior of the gastrointestinal structures. At present AR navigation systems are mostly used for early surgical planning and visualization due to various technical challenges and adoption. One of the challenges in AR integration is superimposing multimodal imaging data onto real-time visuals due to organ movement and deformation during procedures~\cite{gastroenterology}. Recent research has adopted AR technology to improve surgical safety. For example, one study integrated CT imaging data with endoscopic views for skull-base surgery, achieving sub-millimeter accuracy~\cite{Lai2020}, while another combined 3D virtual images with real-time endoscopic footage to enhance visualization in sinus and skull-base procedures~\cite{Li2016}.

Several systems have been redesigned to improve AR navigation visualization in medical data. The MD-cave system \cite{MDcave} supports radiologists with immersive 3D viewing and gesture-based interactions but requires extensive hardware setup. SpatialTouch \cite{spatialTouch} integrated an interactive touch screen for manipulating 3D data with the HoloLens, whereas Haowai et al.~\cite{3dSlicer} integrated the open source 3D Slicer software with AR for medical data visualization. The VRRRRoom system  \cite{VRRRRoom} offers an immersive 3D environment combining HMDs with interactive surfaces, although its limited touchpoints complicate and limit user interactions and exploration.

Advanced rendering techniques have been used to improve the realism and interactivity of 3D medical visualizations. Taibo and Iglesias-Guitian~\cite{StereoscopicVolumetricPath} utilize volumetric path tracing to create high-quality and interactive stereoscopic visualizations of medical data in VR environments. This approach enhances depth perception and realism compared to traditional 2D medical imaging visualizations while allowing real-time navigation and interaction using VR controllers and head tracking. The Medical Visualization Table (MVT) developed by Lundstrom et al.~\cite{Multitouch} provides an intuitive multi-touch interface for visualizing and manipulating 3D medical imaging data for planning orthopedic surgery. It presents 3D renderings of medical images, enabling surgeons to navigate, manipulate, and annotate the data easily while supporting collaborative viewing and interaction. Similarly, Nextmed\cite{gonzalez2020nextmed} offers automated imaging segmentation, 3D reconstruction, and AR-based visualization for enhanced medical imaging analysis. 

Holoview builds on prior AR systems by offering gesture-driven interaction, remote rendering, and real-time stereo visualization, that ensures high-performance, platform-agnostic anatomical exploration. By addressing hardware and interaction limitations, it bridges the gap between clinical and educational AR applications.

\section{Data Preparation and Segmentation}
We use a subset of the Visible Korean Human dataset~\cite{VKH_SukChung2005,VKHanothertrial,VKHanothertrial_1}, which consists of Cryosection images of a male cadaver sectioned at intervals of 0.2 mm, resulting in a resolution of $2468 \times 1407 \times 8506$ of the whole body. We also utilized segmented images of the Cryosection data provided for every fifth anatomical image. To extract relevant information for hybrid rendering while minimizing the dataset's memory footprint, we preprocessed the data as outlined next.

\subsection{Hierarchical Representation of Anatomy}
The Visible Korean Human (VKH) project organizes segmentation data into a three-level hierarchical structure. Level 1 (L$_1$) classifies anatomical structures into 13 major body systems, providing a broad categorization. Level 2 (L$_2$) further refines this classification into individual organs, allowing for more detailed anatomical analysis. Level 3 (L$_3$) captures sub-organ structures, enabling fine-grained examination of complex anatomy.

To encode this hierarchy, each voxel in the segmentation volume is assigned a 16-bit code. The 7 least significant bits represent L$_3$, the next 4 bits correspond to L$_2$, and the following 4 bits encode L$_1$. The most significant bit (MSB) is reserved. A code of zero is assigned to background voxels. Both the client and server share access to the segmentation hierarchy for processing. In this work, we utilize information only up to the L$_2$ level.

\subsection{Correction of mislabeled segments}
The segmentation regions in the original dataset suffer from issues such as missing and incorrect labels in certain slices. To address major problems with the L$_1$ and L$_2$ level segmentation regions, we performed manual inspection and applied semi-automatic techniques. Unlabeled slices were assigned labels based on the annotations of adjacent slices, with manual identification used when neighboring slices had conflicting labels. For cases involving multiple consecutive unlabeled slices, we used a semi-automatic approach, utilizing seeded region growing and level set segmentation methods to delineate the missing class regions.

\subsection{Homotopy based resampling}
The segmentation data is available only for every fifth image of the original Cryosection dataset, which poses the problem of loosing continuity while visualizing the dataset. In order to upsample the segmentation data we use homotopy continuation based approach \cite{sharma2011homotopy} for smooth transition of region boundaries between slices. Homotopy-based resampling ensures smooth and topology-preserving interpolation between segmentation slices that avoids discontinuities common in linear and other interpolation methods. For the same, we formulate a higher-order and Hermite homotopy that provides smooth transition between slices. In our approach, we compute a 2D signed distance field from the segmentation boundaries of each fifth slice and then interpolate this field smoothly for the slices in between. This method is used to upsample the entire segmentation volumes for both  L$_1$ and L$_2$ levels, maintaining continuity throughout the dataset.

\begin{figure}[!h]
\centering
\begin{tikzpicture}
\node[inner sep=0, anchor=south west] (homotopy) at (0, 0){\includegraphics[width=0.75\linewidth]{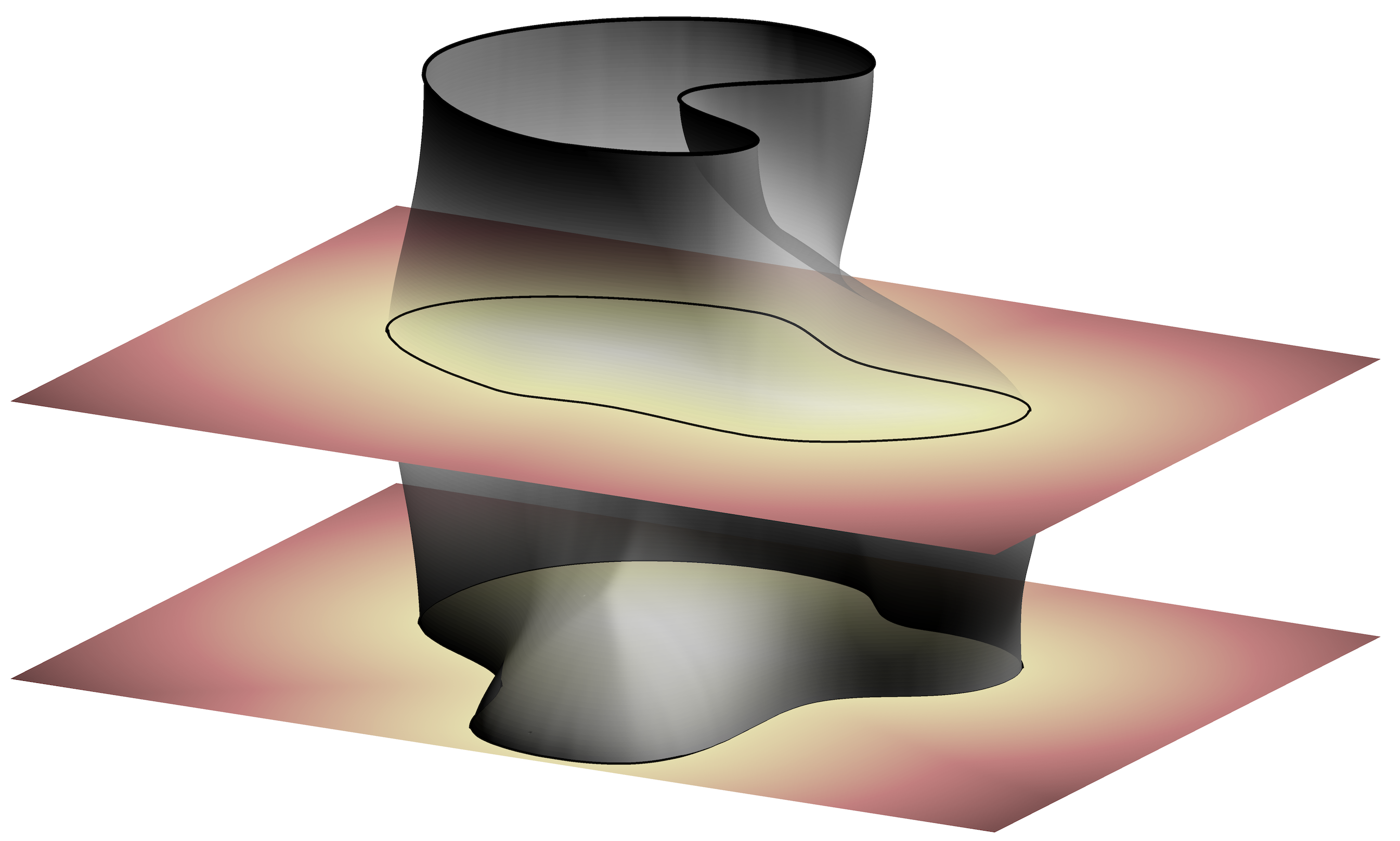}};
\node at (1.75, 3.25) {$\mathcal{S}$};    
\draw[->,-latex] (4.5, 2.5) to [bend left=30] (5.5, 3) node[anchor = west] {$\phi_{i+1}(\mathbf{x})$};  
\draw[->,-latex] (5, 1) to [bend left=30] (5.5, 1.5) node[anchor = west] {$\phi_{i}(\mathbf{x})$};    
\node at (4.5, 3.75) {$s_{i+2}$};    
\node at (5.25, 2) {$s_{i+1}$};    
\node at (4, 0.5) {$s_{i}$};    
\end{tikzpicture}
\caption{Homotopy based resampling.}
\label{fig:homotopy_resampling}
\end{figure}

Given a set of segmentation boundaries $\{s_i\}$ for slices $i\in\{1,\ldots, n\}$, we compute the signed distance functions $\phi_i(\mathbf{x})$, which represent the signed distance from any point $\mathbf{x}$ to the corresponding boundary $s_i$. A piecewise cubic Hermite homotopy is formulated as
\begin{align}
\mathcal{H}_i(\mathbf{x}, \lambda) = \frac{1}{2} \Big(&\left(2-\lambda-4\lambda^2 + 3\lambda^3\right)\phi_i(\mathbf{x})\nonumber \\
+&\left(\lambda + 5\lambda^2-4\lambda^3\right)\phi_{i+1}(\mathbf{x})\nonumber \\
+&\left(-\lambda^2 + \lambda^3\right)\phi_{i+2}(\mathbf{x})\Big), \quad\textrm{for\ } \lambda\in[0,1],
\end{align}
which is derived by setting derivatives at the boundaries of $\mathcal{H}_i$ to 
\begin{align}
\frac{\partial\mathcal{H}_i(\mathbf{x}, \lambda)}{\partial\lambda}\bigg|_{\lambda = 0} &= \frac{1}{2} \big( \phi_{i+1}(\mathbf{x}) - \phi_{i}(\mathbf{x}) \big), \\  
\frac{\partial\mathcal{H}_i(\mathbf{x}, \lambda)}{\partial\lambda}\bigg|_{\lambda = 1} &= \frac{1}{2} \big( \phi_{i+2}(\mathbf{x}) - \phi_{i+1}(\mathbf{x}) \big).
\end{align}  

The final surface $\mathcal{S}$ is defined as the zero level set of the combined homotopies, given by  
\begin{align}
\mathcal{S} = \{ \mathbf{x} \mid \exists i, \exists\lambda\in[0, 1] \text{ such that } \mathcal{H}_i(\mathbf{x}, \lambda) = 0\}.
\end{align}
Using homotopy-based resampling, we achieve smooth interpolation between intermediate segmentation slices. The resampled segmentation grids are then utilized to accurately determine opacities, ensuring proper alignment with organ structures during volume visualization. For use with our AR framework, we downsampled the Cryosection and resampled segmentation voxel grids to half of their original dimensions, achieving a resolution of $1234 \times 703 \times 4253$.

\subsection{Mesh Generation for Interactive Rendering}
Surface meshes are extracted from the resampled segmentation grids (for both L$_1$ and L$_2$ levels) and utilized to accelerate our ray-marching-based rendering process. By using these meshes, we optimize rendering by enabling empty space skipping along rays, significantly improving computational efficiency.

Starting with surface meshes extracted using the marching cubes algorithm, we applied decimation to reduce triangle count while preserving overall shape. To address the roughness introduced by decimation, we used the Taubin smoothing filter \cite{taubinFilter}, which reduces surface irregularities without significant shrinkage. Despite smoothing, high triangle counts remained a concern for efficient rendering. To optimize further, we applied the adaptive remeshing filter of Attene  and Falcidieno \cite{remeshing}, which reduces triangle density while preserving essential details, ensuring efficient and accurate surface details. 

\section{System Design and Implementation}

\begin{figure*}[!h]
    \centering
    \begin{tikzpicture}
    	\node[fill=gray!50,thick,rounded corners=.2cm,minimum size=0.6cm,inner sep=0.2cm, text width=3.5cm] at (4,2) (A) {\textsc{Client (HoloLens)}\ \faIcon{glasses}};
    	\node[draw,rounded corners=.2cm, inner xsep=0 cm, inner ysep=0.15cm,  anchor=north] at ($(A.south) + (0, -0.5)$) (A1)
			{\begin{tabular}{ccc}
			\Huge \faIcon{vr-cardboard} & \Huge \faIcon{hand-sparkles} & \Huge \faIcon{eye}\\
			\small \emph{Head} & \small \emph{Hand} & \small \emph{Eye}
			\end{tabular}};
		\node[fill=white, inner sep=0, anchor=center] at (A1.north) {\small \bf{Tracking}};
		
    	\node[fill=gray!50,thick,rounded corners=.2cm,minimum size=0.6cm,inner sep=0.2cm, text width=3.5cm] at (12,2) (B) {\textsc{Server (Renderer)}\ \faIcon{server}};
    	\node[draw,rounded corners=.2cm, inner xsep=0 cm, inner ysep=0.15cm,  anchor=north] at ($(B.south) + (0, -0.5)$) (B1)
			{\begin{tabular}{ccc}
			\Huge \faIcon{images} & \Huge \faIcon{cubes} & \Huge \faIcon{dot-circle}\\
			\small \emph{Stereo} & \small \emph{Hybrid} & \small \emph{Foveated}
			\end{tabular}};
		\node[fill=white, inner sep=0, anchor=center] at (B1.north) {\small \bf{Frame rendering}};
	
	 	\draw[thick,black,->,arrows = {-Stealth[harpoon]}] ($(A.east) + (0.1,0.1)$) to node[above] {\small \emph{Control Stream}} ($(B.west) + (-0.1,0.1)$);  
	 	\draw[thick,black,->,arrows = {-Stealth[harpoon]}] ($(B.west) + (-0.1,-0.1)$) to node[below] {\small \emph{Data Stream}} ($(A.east) + (0.1,-0.1)$);  
		
    \draw [decorate,decoration={calligraphic brace,mirror,amplitude=10pt}, very thick] 
        ($(A1.west) + (0, -1)$) -- ($(B1.east) + (0, -1)$) node[midway,below=15pt] {};  
        \matrix[row sep=0.1cm, column sep=0.5cm] at ($(A1)!0.5!(B1) + (0,-3.5)$) {        
        \node (n1) [circle, draw, gray!50, thick, minimum size=3.5cm, inner sep=0pt, 
            path picture={\node at (path picture bounding box.center)
            {\includegraphics[width=3cm]{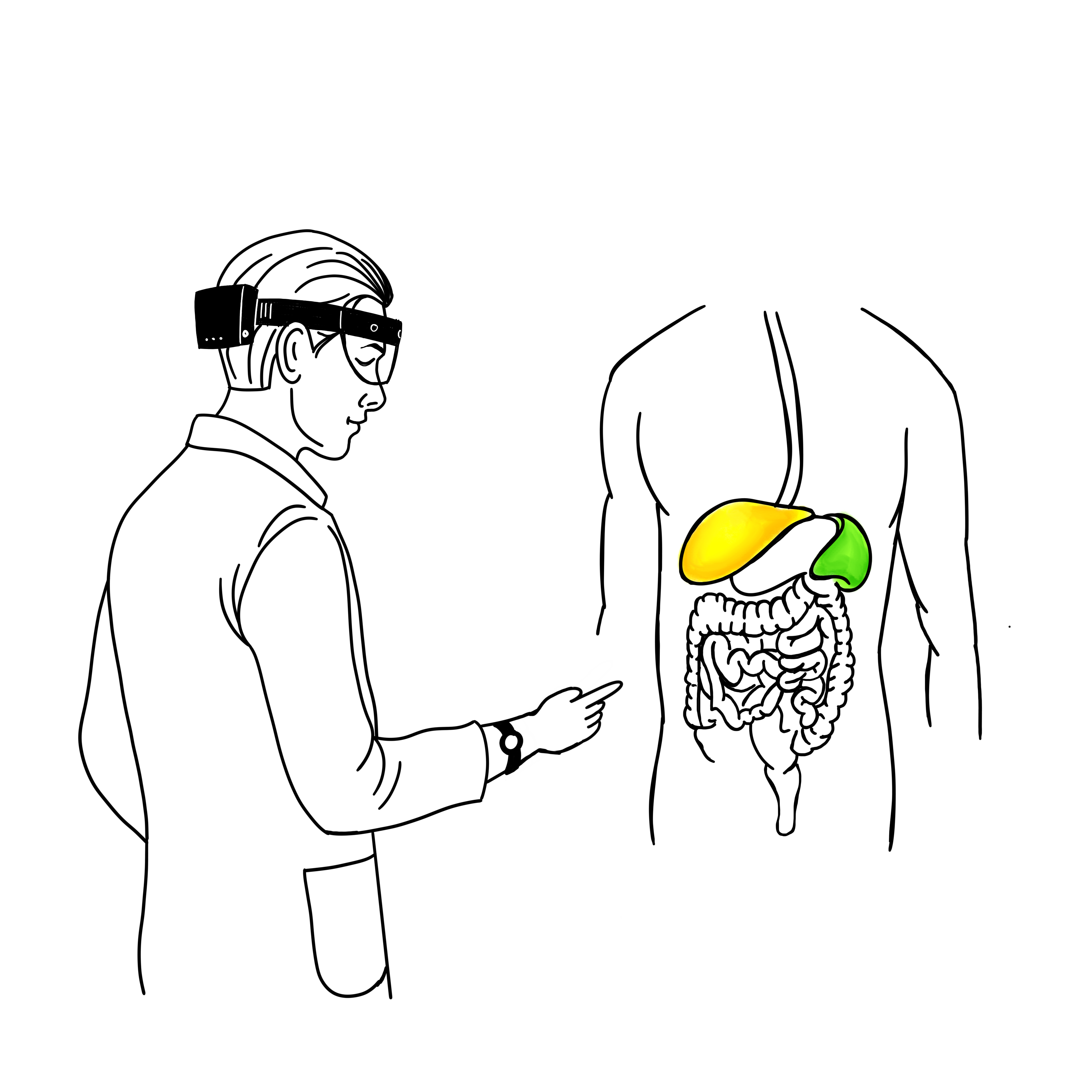}};}] {}; &
        \node (n2) [circle, draw, gray!50, thick, minimum size=3.5cm, inner sep=0pt, 
            path picture={\node at (path picture bounding box.center)
            {\includegraphics[width=3cm]{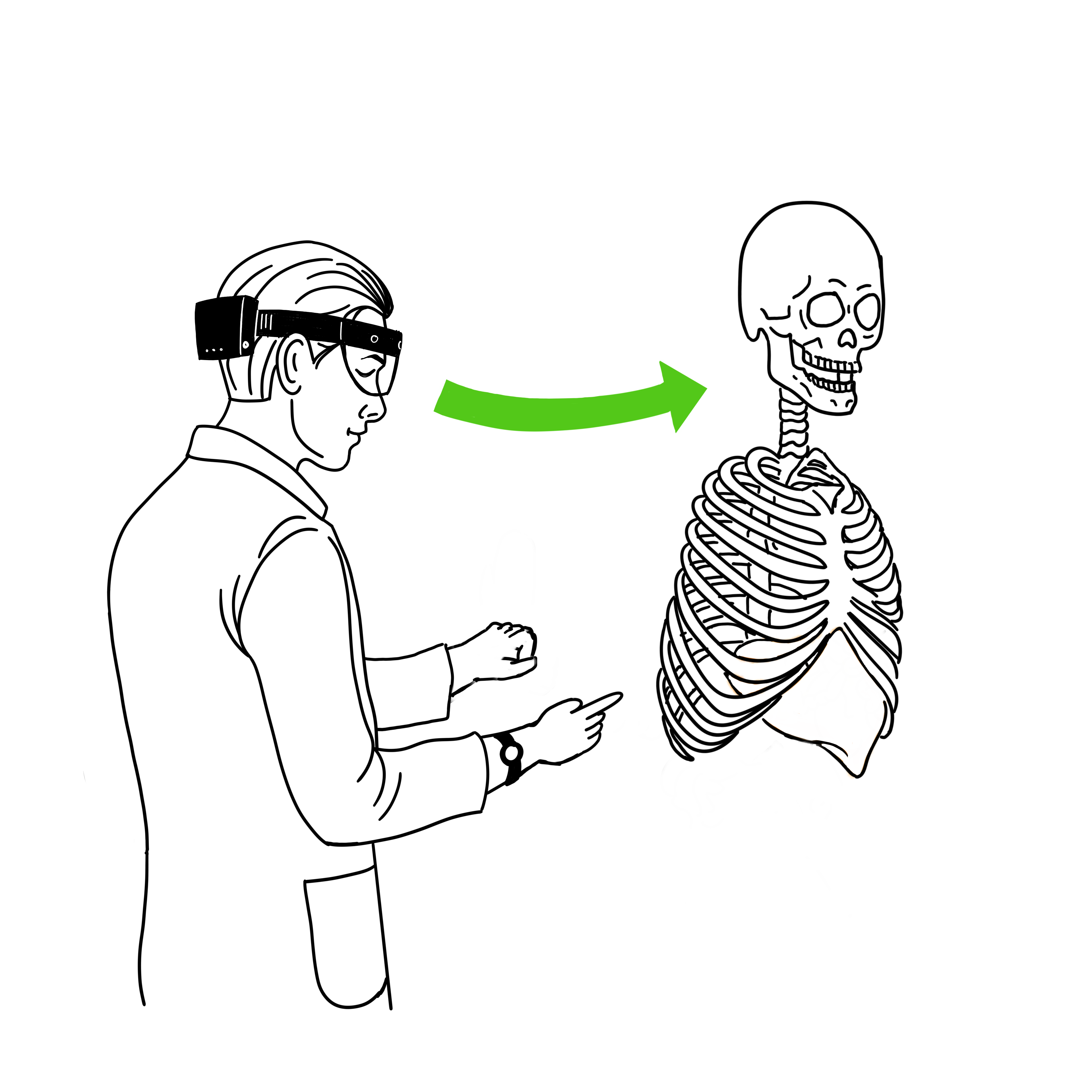}};}] {}; &
        \node (n3) [circle, draw, gray!50, thick, minimum size=3.5cm, inner sep=0pt, 
            path picture={\node at (path picture bounding box.center)
            {\includegraphics[width=3cm]{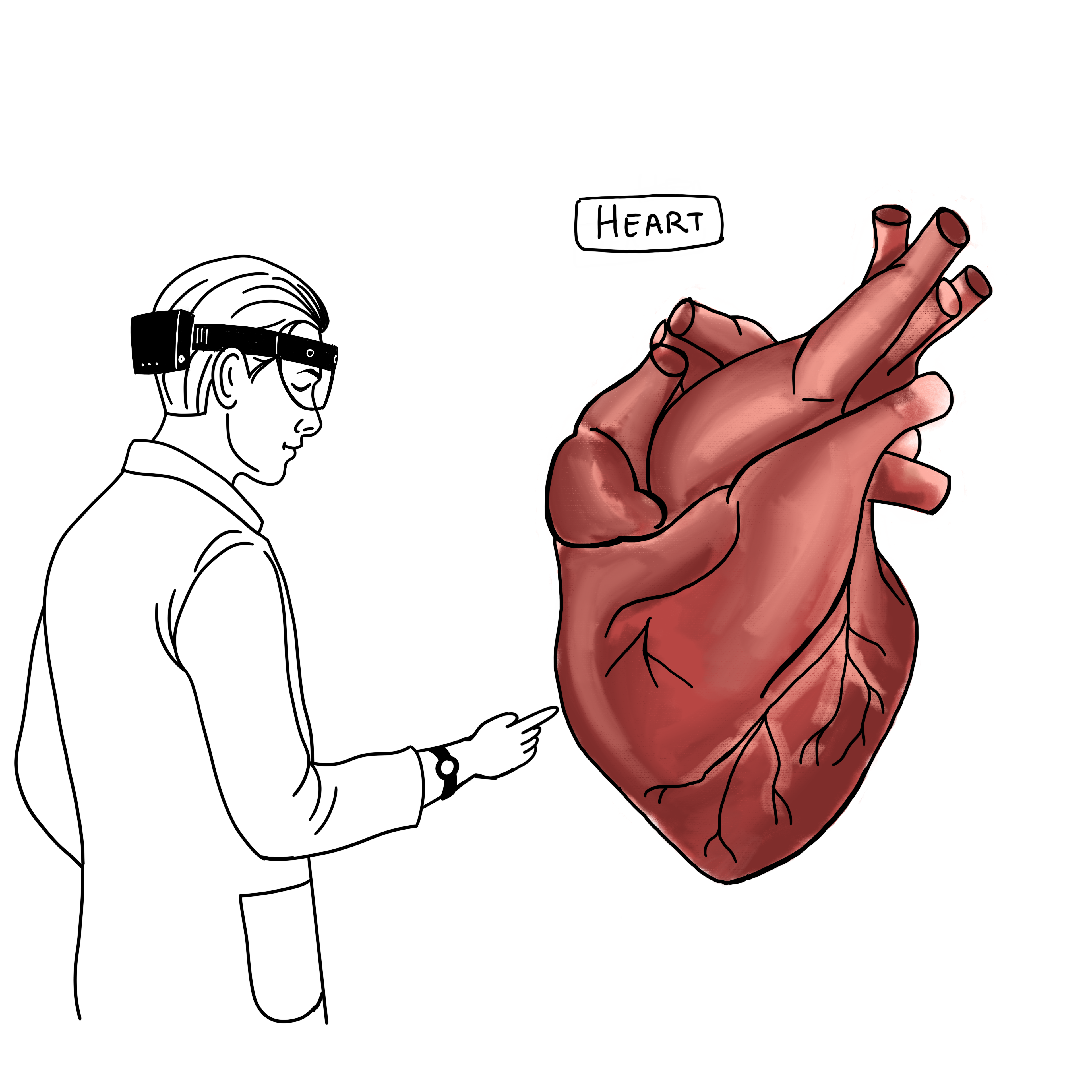}};}] {}; &
        \node (n4) [circle, draw, gray!50, thick, minimum size=3.5cm, inner sep=0pt, 
            path picture={\node at (path picture bounding box.center)
            {\includegraphics[width=3cm]{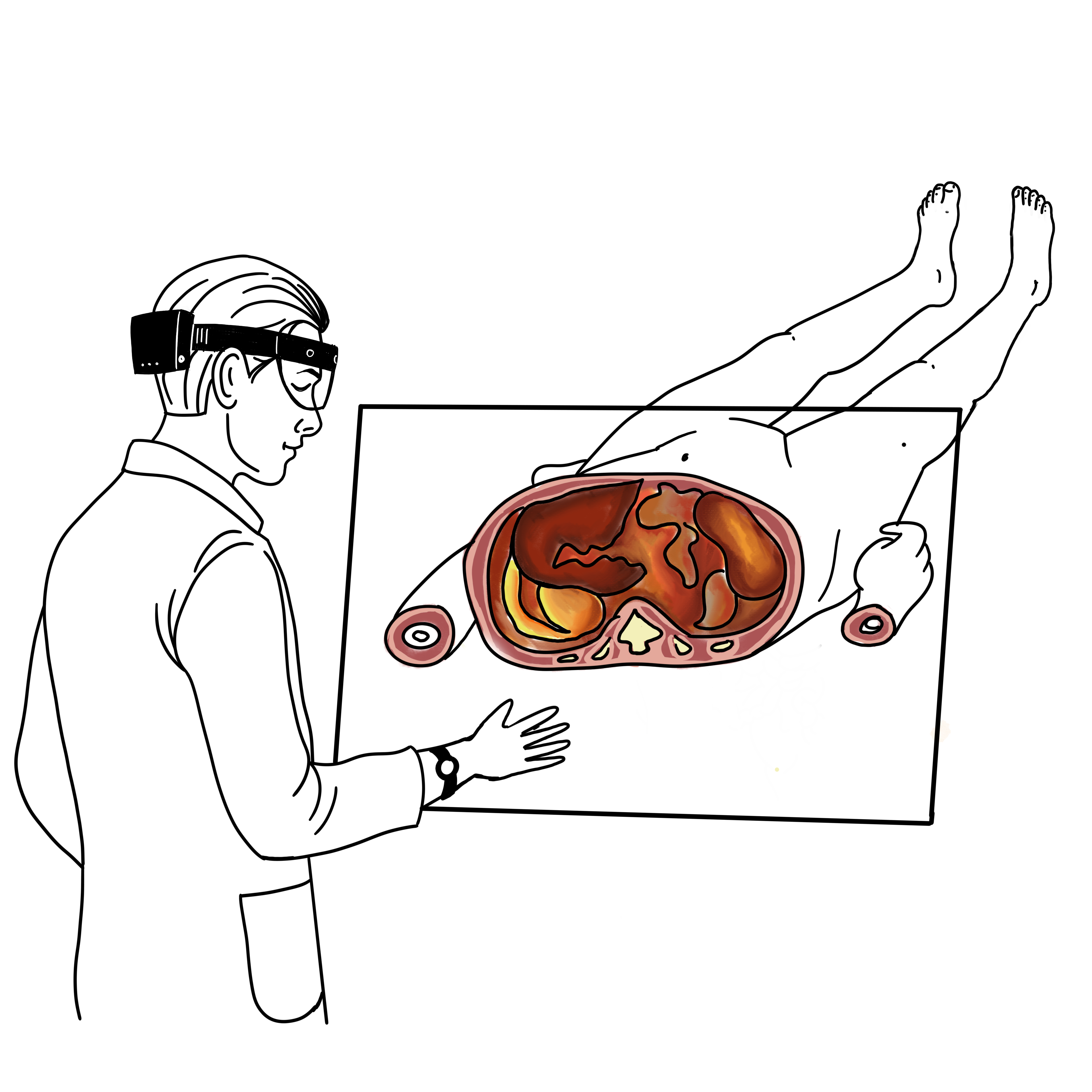}};}] {}; \\
        \node {\small \emph{Selective Rendering}}; &
        \node {\small \emph{Navigation}}; &
        \node {\small \emph{Bioscope Mode}}; &
        \node {\small \emph{Clipping Plane}}; \\
    	};       
    \end{tikzpicture}
    \caption{Overview of the Holoview system architecture. The Hololens client transmits tracked user input to a remote high-performance rendering server. Holoview supports various interaction modes: selective rendering of anatomical structures, navigation within volumetric data, bioscope for detailed organ inspection, and a clipping plane for cross-sectional viewing.}
    \label{fig:overview} 
\end{figure*}

Holoview is an Augmented Reality (AR) system designed to support anatomical education through an immersive and intuitive interface. It is based on a multi-threaded remote rendering architecture, wherein real-time stereo rendering of large volumetric datasets is performed on a high-performance server. The Microsoft HoloLens functions as a lightweight thin client, responsible for display and user tracking. Head, hand, and eye-gaze orientations are tracked in real time to enable natural interaction. User inputs, including hand gestures, camera orientation, and rendering parameters, are transmitted over the network via a \emph{control stream} to the remote renderer. Based on this data, the server generates stereo frames, encodes them, and transmits them back via a \emph{data stream} for real-time display on the AR device. This architecture ensures seamless interactivity without the need for external controllers.

Holoview enables selective visualization of anatomical structures and contextual information through intuitive gestures. A dual-hand navigation mechanism allows movement within confined spaces: the left hand activates navigation mode, while the right hand controls direction. Internal anatomical exploration is facilitated through dynamic tissue visibility adjustments and a virtual clipping tool, which attaches a cutting plane to the user's right hand for precise cross-sectional views. Figure~\ref{fig:overview} provides an overview of the system architecture.

The client application is developed in C and C\# within the Unity game engine, utilizing the Microsoft Mixed Reality Toolkit 3 (MRTK) for HoloLens integration. The server is implemented in C++ using CUDA and OptiX \cite{optix}, running on a Quadro workstation with an Nvidia RTX A5000 GPU. Stereo frame rendering for both eyes is executed in parallel using CUDA streams, with results sent via a separate data stream thread.

\subsection{Hybrid rendering}
\label{ssec:hybrid_rendering}
\begin{figure}[!h]
\centering
\begin{tikzpicture}
\node[inner sep=0, anchor=south west] (img) at (0, 0){\includegraphics[width=0.9\linewidth]{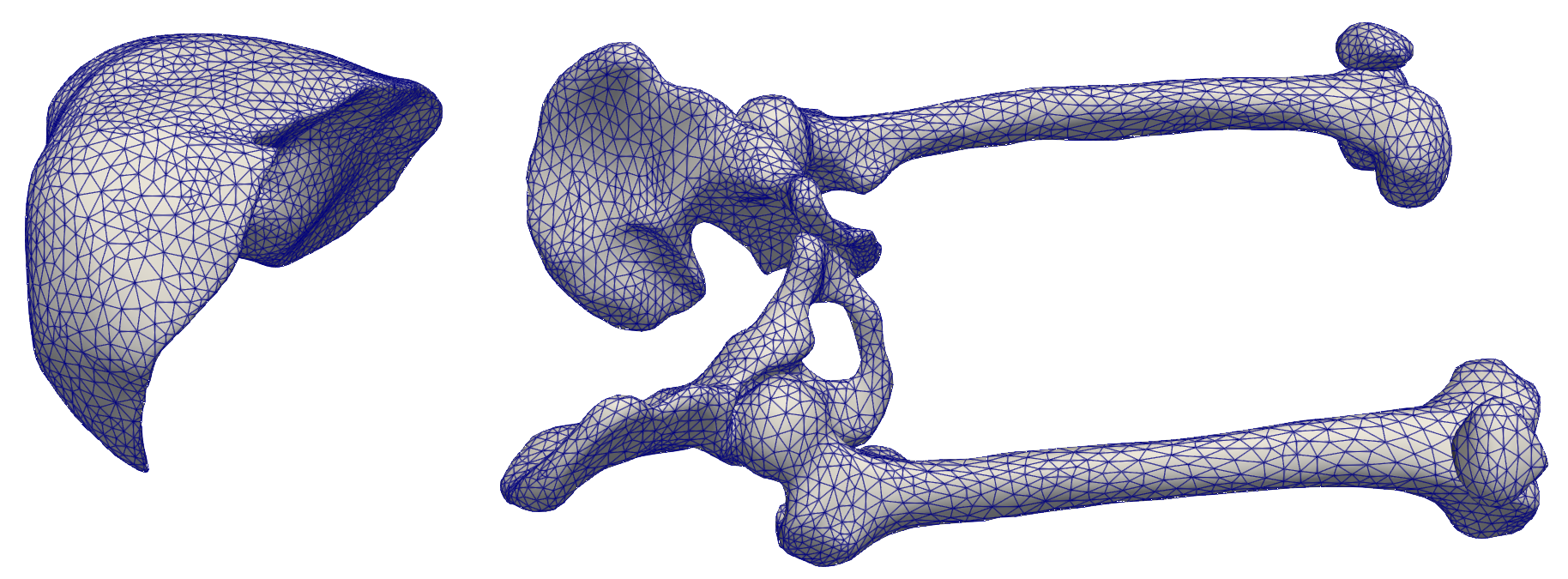}};
\node[rotate=20, anchor=center] (cam) at (-0.5,1) {\Large \faIcon{video}};

\draw[thick, rotate=20, anchor=center, teal] (0.2, 1.1) -- (0.75, 1.1);
\draw[thick, rotate=20, anchor=center, teal, opacity=0.5] (1.5, 1.1) -- (1.68, 1.1);

\draw[very thick, rotate=20, anchor=center, yellow, line cap=round, opacity=0.5] (0.75, 1.1) -- (1.5, 1.1);
\draw[thick, rotate=20, anchor=center, red, dash pattern=on 0pt off 2\pgflinewidth, line cap=round, opacity=0.75] (0.75, 1.1) -- (1.5, 1.1);

\draw[thick, rotate=20, anchor=center, teal] (1.68, 1.1) -- (3.5, 1.1);

\draw[thick, rotate=20, anchor=center, teal, opacity=0.5] (4, 1.1) -- (4.21, 1.1);

\draw[very thick, rotate=20, anchor=center, yellow, line cap=round, opacity=0.5] (3.5, 1.1) -- (4, 1.1);
\draw[thick, rotate=20, anchor=center, red, dash pattern=on 0pt off 2\pgflinewidth, line cap=round, opacity=0.75] (3.5, 1.1) -- (4, 1.1);

\draw[thick, rotate=20, anchor=center, teal, -{Latex[length=2mm, width=1mm]}] (4.21, 1.1) -- (4.75, 1.1);

\fill[rotate=20] (0.75, 1.1) circle (1pt);
\fill[rotate=20] (1.5, 1.1) circle (1pt);
\fill[rotate=20] (3.5, 1.1) circle (1pt);
\fill[rotate=20] (4, 1.1) circle (1pt);

\node at (0.25, 3) (txtxmesh) {\small \emph{Mesh intersection}};
\draw[rotate=20, ->, -latex] (0.7, 1.2) to [bend left=30] (txtxmesh.south);
 
\node at (2, 1) (txtempty) {\small \emph{Empty space}};
\draw[rotate=20, ->, -latex] (2.5, 1) to [bend left=15] (txtempty.north);

\end{tikzpicture}
\caption{Mesh-guided volume rendering. Organ meshes define ray entry and exit points, enabling selective visualization during ray marching.}
\label{fig:hybrid_rendering}
\end{figure}

Holoview incorporates selective rendering (see Subsection~\ref{ssec:selective_rendering}) that enables users to interactively choose and visualize specific organs. This process dynamically updates the alpha values of all voxels, however setting non-target structures to zero opacity introduces large empty regions, which significantly reduces frame rates. To address this issue, Holoview incorporates mesh-based empty space skipping in the volumetric grid. Meshes for L$_2$ organs serve as surface-based approximations and are used to precisely define pairs of entry and exit points along each ray for color accumulation (see Figure~\ref{fig:hybrid_rendering}). Ray traversal is performed only within the selected organs, thus eliminating unnecessary computations leading to better rendering efficiency and higher frame rates. This hybrid approach maintains a balance between the detailed representation of volumetric data and the computational efficiency of surface rendering, offering an optimized solution for interactive medical visualization.

\subsection{Network optimization}
To ensure optimal network performance for our augmented reality (AR) system, we established a local network using a Wi-Fi router as a dedicated access point. The server was connected to the router via ethernet, while the HoloLens connected wirelessly. We utilized Wi-Fi 6 (802.11ax) with a 160 MHz band to achieve high-speed, low-latency communication essential for real-time rendering and interaction. Additionally, we optimized the wireless environment by enabling beamforming, minimizing physical obstructions, and selecting an efficient communication channel.

\subsection{Foveated rendering}
The HoloLens 2 has a native resolution of $1440\times 960$, but transferring rendered stereo images at this resolution results in a frame rate of \textasciitilde{}7fps. In an eye-tracked AR setup, display performance can be improved through foveated rendering, which reduces the number of pixels to be rendered away from the gaze point. To address this, we implemented the rectangular mapping-based foveated rendering (RMFR) method by Ye et al.~\cite{foveated}. RMFR adapts the levels of foveation based on the scene complexity and provides high quality reconstructions even with a high compression of ~15\% shading samples. We rendered stereo images at one-third of the resolution in each dimension while maintaining high quality near the gaze point. This approach increases the image transmission rate to 60\textasciitilde{}70 fps while preserving visual quality for the user.

\section{Interaction Design and Features}

Designing effective gesture-based interactions for mixed reality systems like Holoview requires a systematic approach that balances intuitiveness, precision, and ergonomic comfort. Since HoloLens 2 relies entirely on hand gestures for interaction, we developed a dedicated gesture interaction pipeline that ensures robustness, learnability, and minimal cognitive load during use.
 
\subsection{Interaction Design}

\paragraph{Design Process Overview}
The interaction design for Holoview focused on two critical components—gesture controls and the spatial UI system. Primary interactions identified included selecting anatomical structures, navigating within the cadaver, activating tools like the clipping plane, retrieving anatomical labels, switching to focused exploration modes (such as Bioscope), and manipulating models via translation, rotation, and scaling. Each function was categorized based on frequency of use, required precision, and spatial complexity. For example, selection needed to be extremely fast and accurate, while mode-switching could be slower but demanded deliberate gestures to avoid accidental activations. Similarly, UI design had to minimize visual clutter, reduce hand fatigue, and ensure alignment with these task flows without breaking immersion.

\paragraph{Ideation, Prototyping, and Testing}  
We explored a broad range of gesture possibilities, inspired by natural hand movements, existing AR/VR interaction taxonomies, and prior research. More than 30 single-handed and two-handed candidate gestures were proposed and prototyped using Unity and MRTK. In parallel, we experimented with spatial UI layouts, ensuring that menus and controls were anchored relative to the anatomy, not the user’s gaze. 

To evaluate these designs, we conducted two usability testing rounds with six participants, including both MR novices and experienced users. They were asked to complete anatomical inspection tasks using different UI and gesture variants. Metrics such as task completion time, gesture recognition, interface clarity, hand fatigue, and user preference were used to define the best UI and gestures.

\paragraph{Iterative Refinement}  
Test results drove the refinement process. For gestures, complex finger sequences were simplified, misrecognized gestures were discarded, and all overlaps were resolved. For UI, spacing, contrast, and proximity to key anatomy were adjusted for comfort and visibility. The final interaction set was selected based on high recognition accuracy, low physical and cognitive load, ease of learning, and alignment with user expectations. Together, gesture vocabulary and UI system were tightly integrated to support Holoview’s embodied, controller-free design philosophy.

\subsubsection{Gesture Design}
The gesture system in \textit{Holoview} was crafted to be natural, reliable, and efficient for core interaction tasks in a controller-free environment. Final gestures were selected to balance ergonomics, ease of learning, and intentionality—enabling users to focus on anatomical content rather than interface complexity.

To ensure accurate recognition, we implemented spatial and temporal constraints using MRTK. Gestures like Bioscope activation required users to hold a specific pose for at least 500 milliseconds, reducing accidental triggers. Bimanual gestures were recognized only when hands met defined spatial proximity and orientation thresholds, ensuring clear differentiation between gestures.

Multimodal feedback was integrated to reinforce interaction flow. Visual cues—such as glowing UI elements and highlighting anatomical regions—were paired with motion-based confirmations, including animations like smooth zoom-in transitions. This layered response model provided immediate, context-aware feedback, improving gesture clarity and reducing uncertainty.

All gestures adhered to key design principles: they were consistent, intentional, and mapped to familiar physical actions. Critical operations used deliberate poses to maintain control, while consistency across gesture types supported rapid learning and mental model formation. The final gesture set allowed seamless transitions across exploration, selection, and manipulation tasks without requiring mode shifts or cognitive reorientation—crucial for maintaining user focus in a medical learning context.

\subsubsection{UI Design}
The user interface of \textit{Holoview} is a key enabler of intuitive interaction in a controller-free, gesture-based mixed reality environment. Entirely designed in world space, the UI components are spatially anchored relative to the anatomical model rather than following the user’s gaze. This spatial consistency enhances focus, reduces visual clutter, and maintains immersion during anatomical exploration.

\begin{figure}
    \centering
    \begin{tabular}{ccc}
        \includegraphics[width=0.3\linewidth]{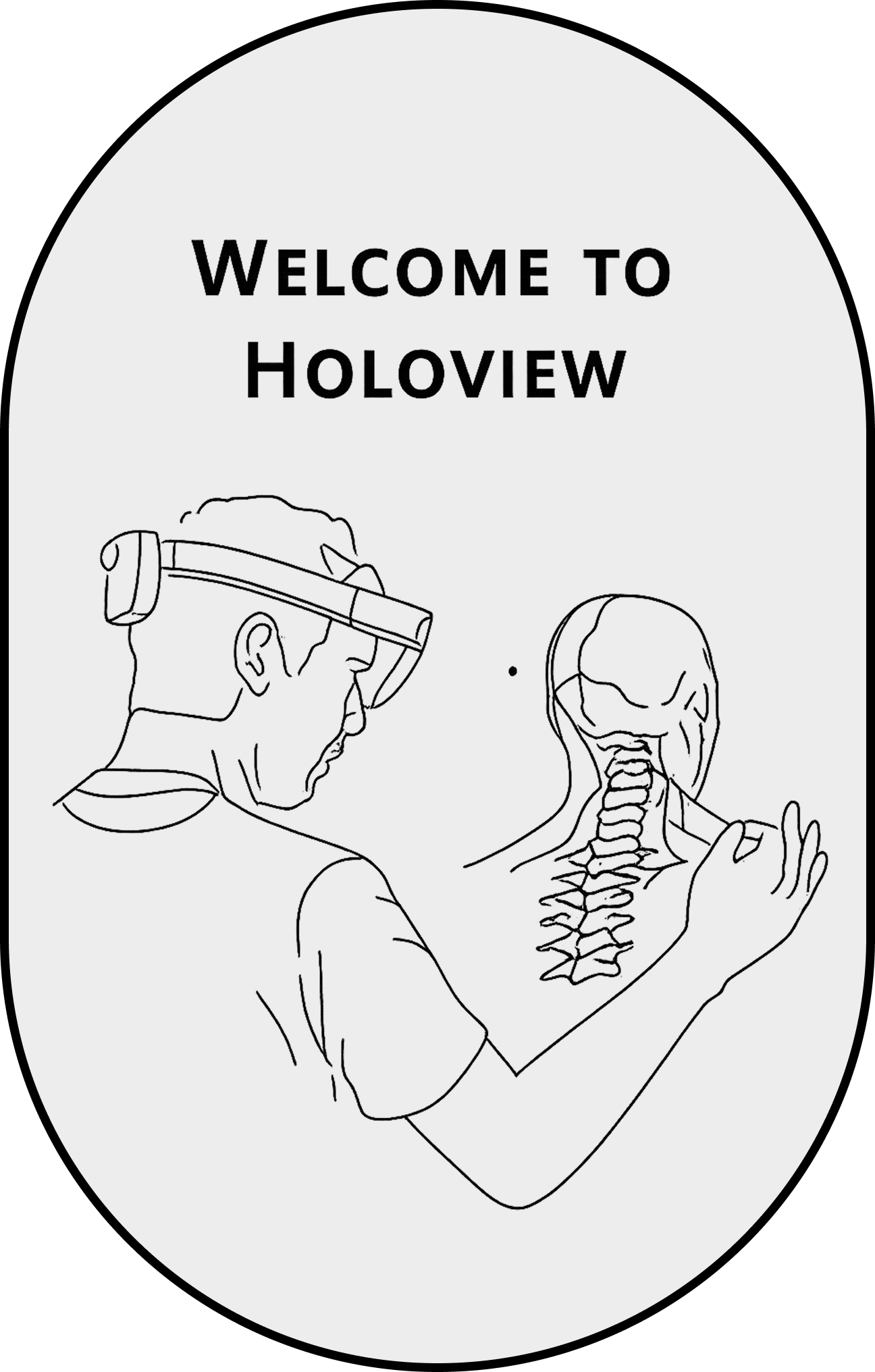} &
        \includegraphics[width=0.3\linewidth]{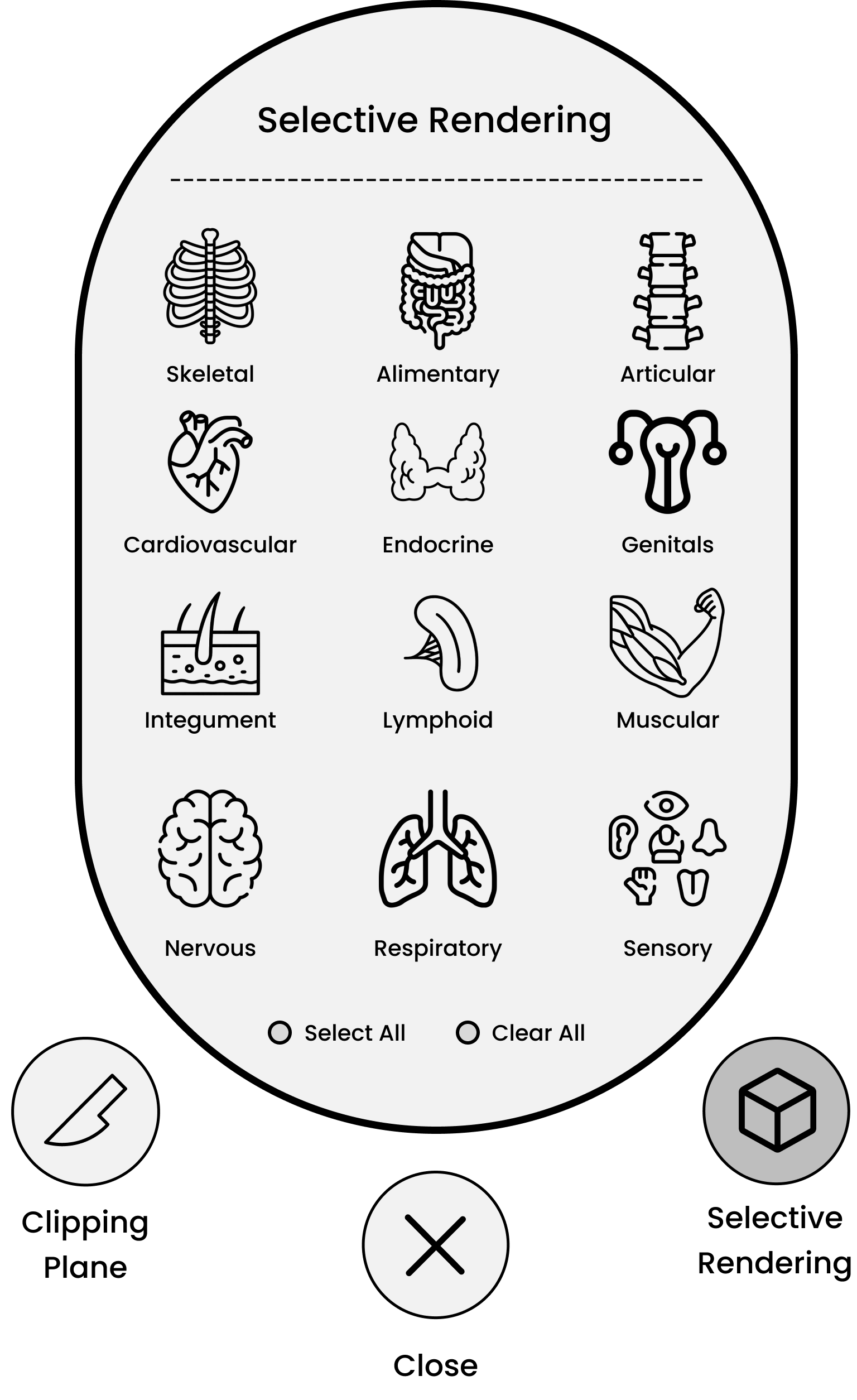} &
        \includegraphics[width=0.3\linewidth]{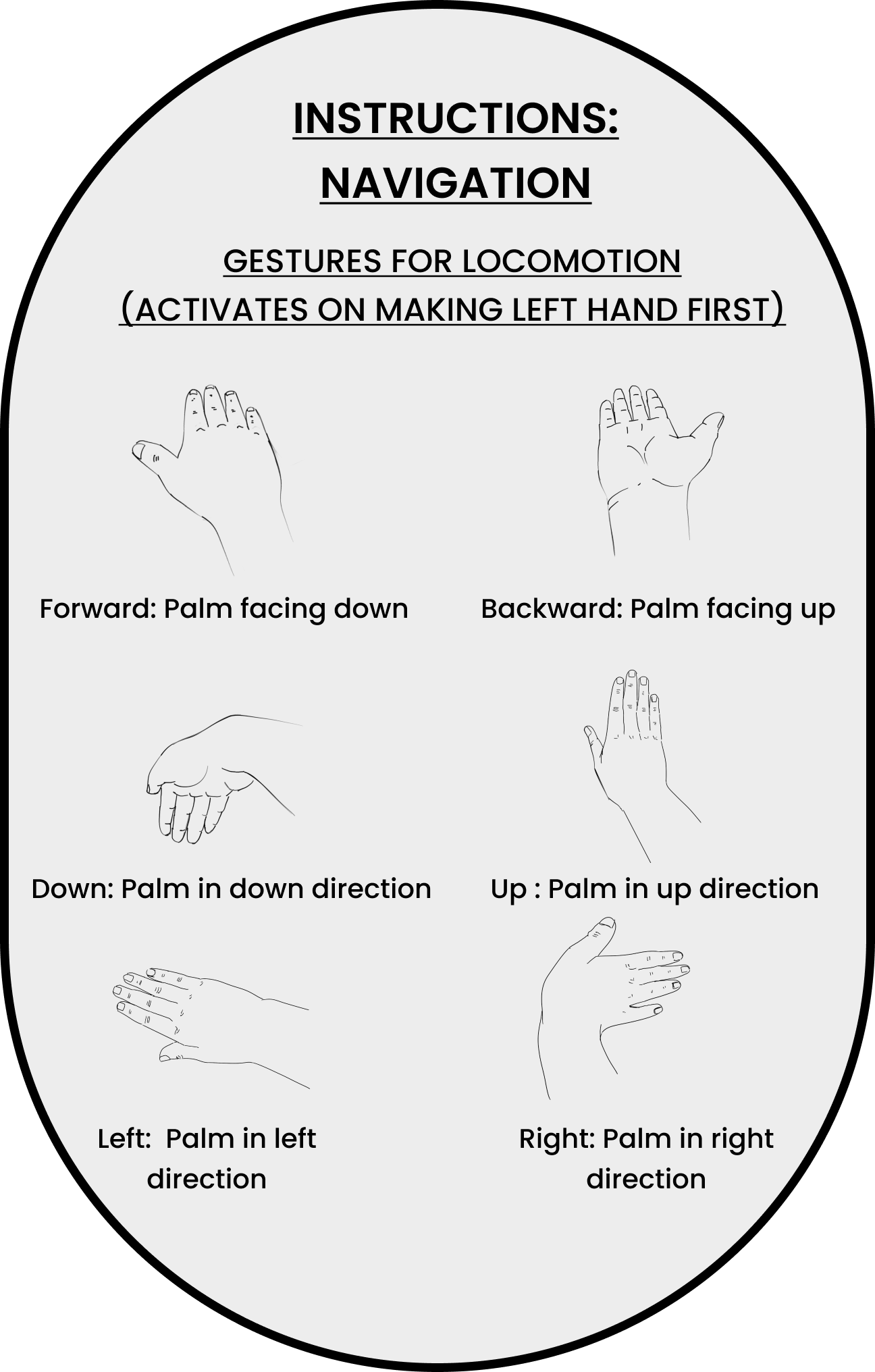} \\
        (a) & (b) & (c)
    \end{tabular}
    \caption{Holoview interactive UI. (a) Welcome screen, (b) Organ selection UI, (c) Instructions UI.}
    \label{fig:UI_final}
\end{figure}

UI elements are positioned contextually within the AR environment to encourage natural interaction. For instance, the welcome screen (Figure~\ref{fig:UI_final}(a)) offers access to core functions such as the clipping plane, selective rendering, and exit in a compact layout that facilitates orientation without overwhelming the user. The organ selection panel (Figure~\ref{fig:UI_final}(b)) appears adjacent to the cadaver, displaying L$_1$ organ systems as intuitive icons that respond to hand proximity or pointer gestures with highlights or glow effects.

Users can summon the UI at any time by raising their left hand with the palm facing upward. The menu then appears above the palm, providing a physically grounded access point. Once the hand is lowered, the menu remains fixed in place. Repeating the gesture relocates the menu, offering spatial flexibility and reinforcing Holoview’s embodied interaction model. This approach minimizes hand fatigue and enables task-aligned access to controls.

To support discoverability and ease of use, Holoview includes a dynamic instructional panel attached to the side of the main UI. This context-sensitive guide updates automatically with gestures, labels, and icons relevant to the active mode—such as clipping or navigation—providing in-situ assistance. By embedding guidance into the interface, the system reduces learning overhead and promotes fluid task progression without interrupting the user experience.

\subsection{Features}
\begin{figure*}
    \centering
    \begin{tabular}{cccc}
    \includegraphics[height=0.18\textheight]{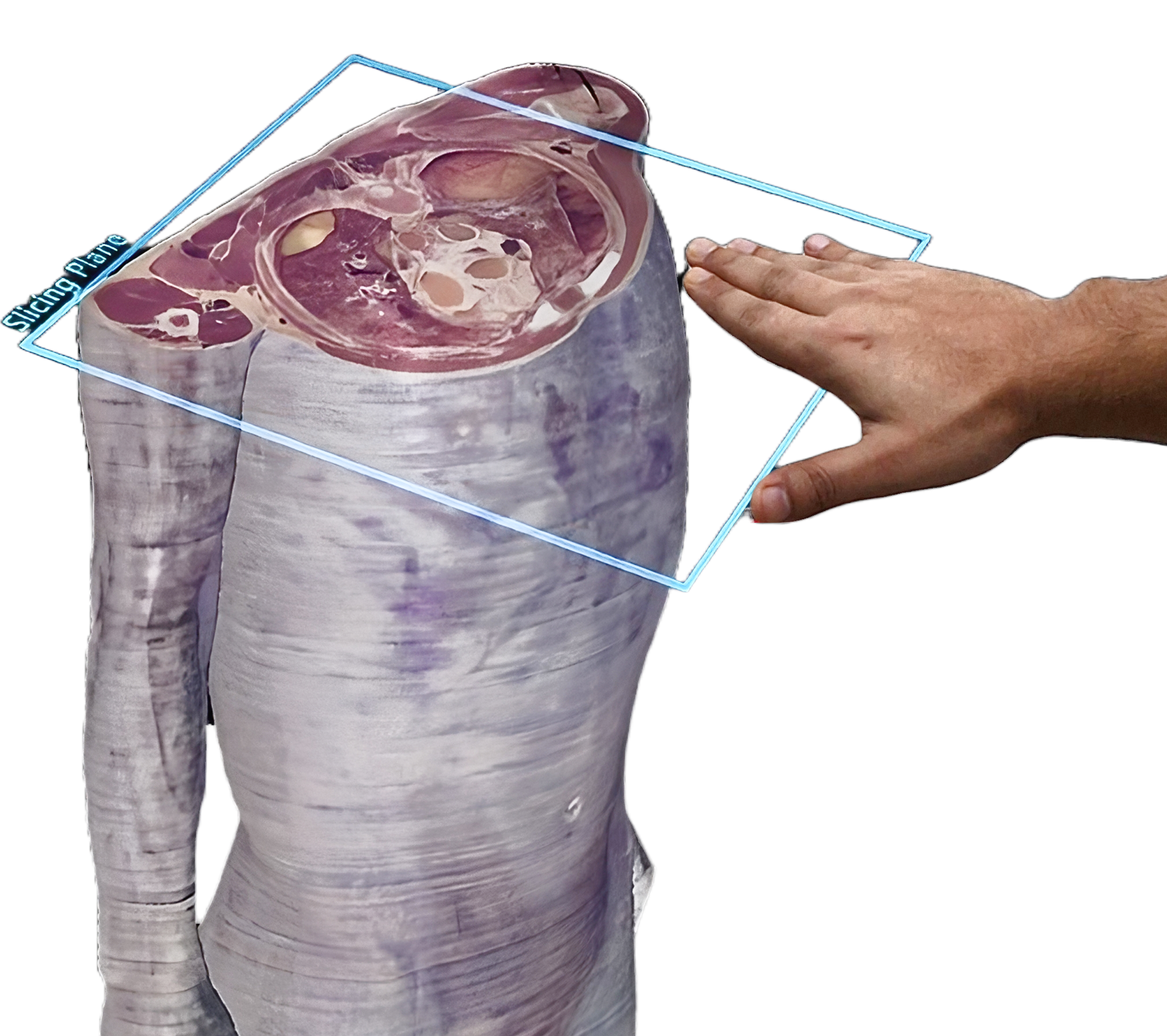} & 
    \includegraphics[height=0.18\textheight]{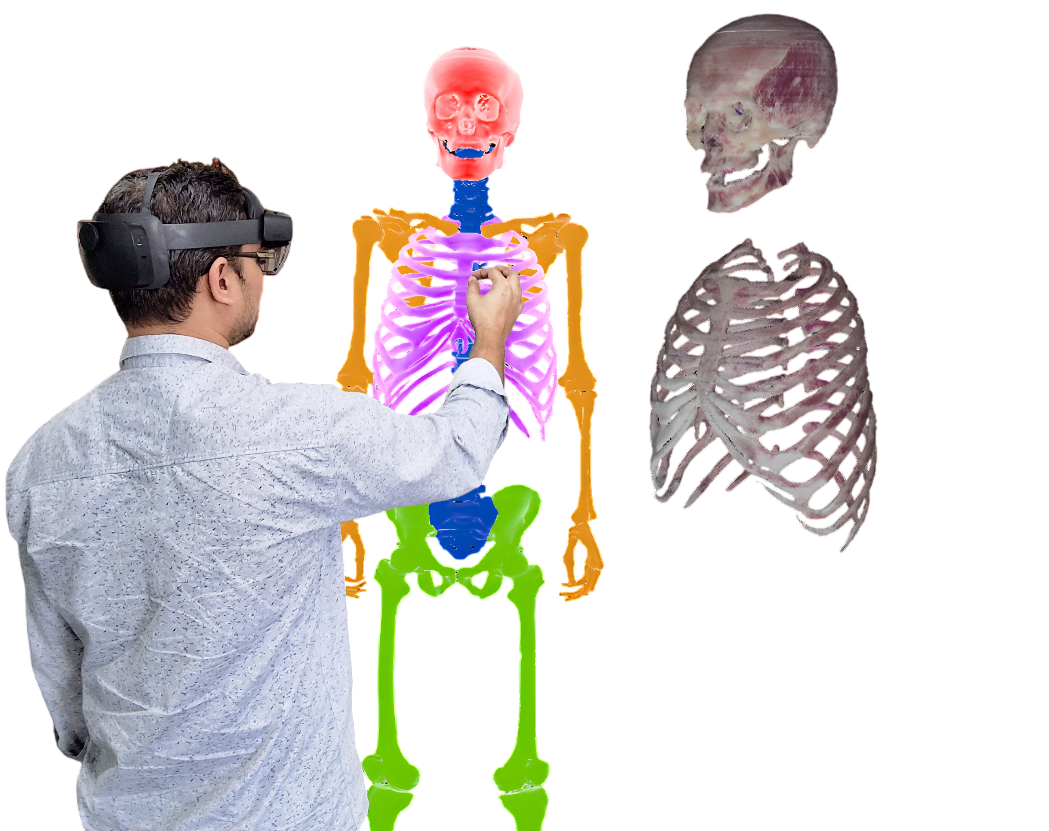} & 
    \includegraphics[height=0.18\textheight]{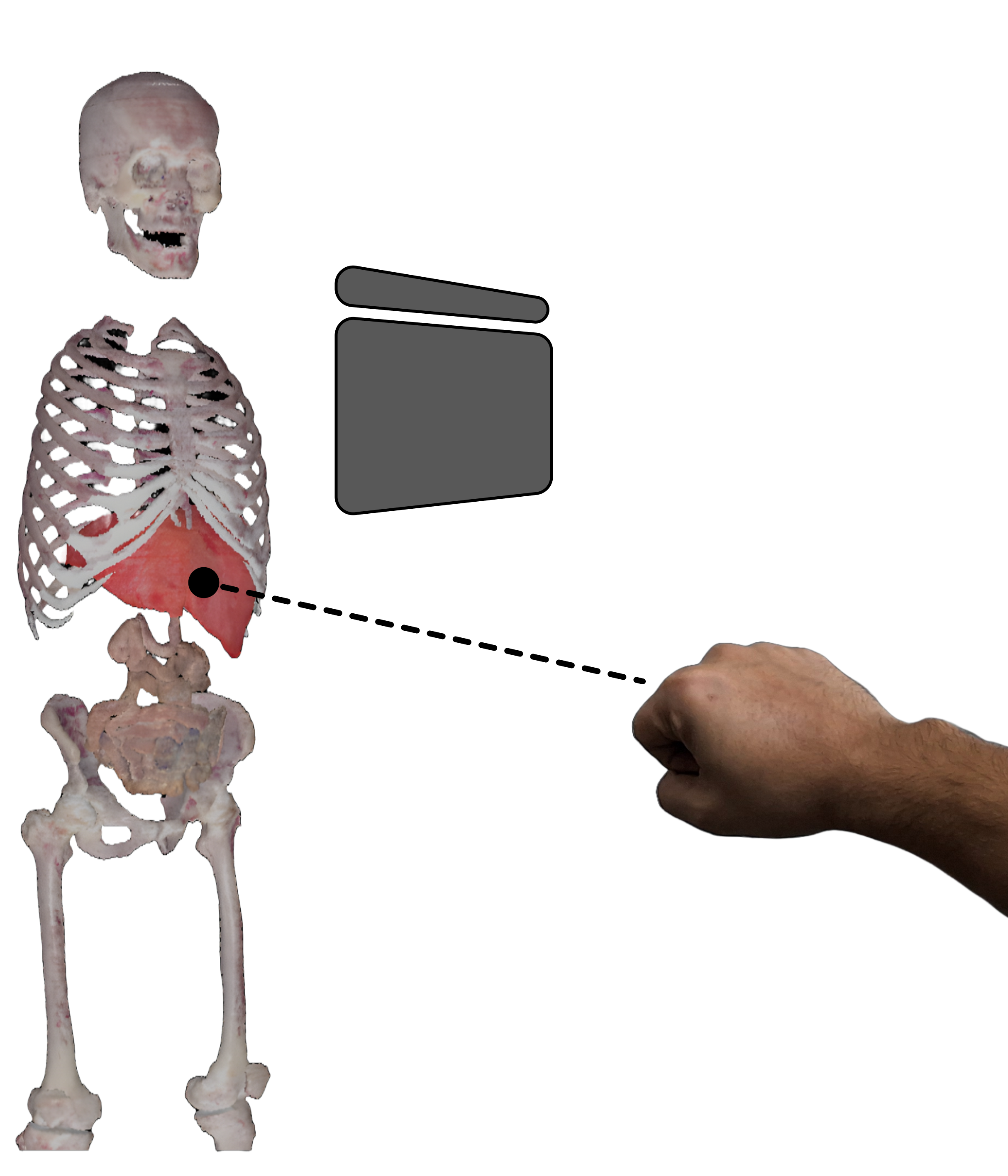} &
    \includegraphics[height=0.18\textheight]{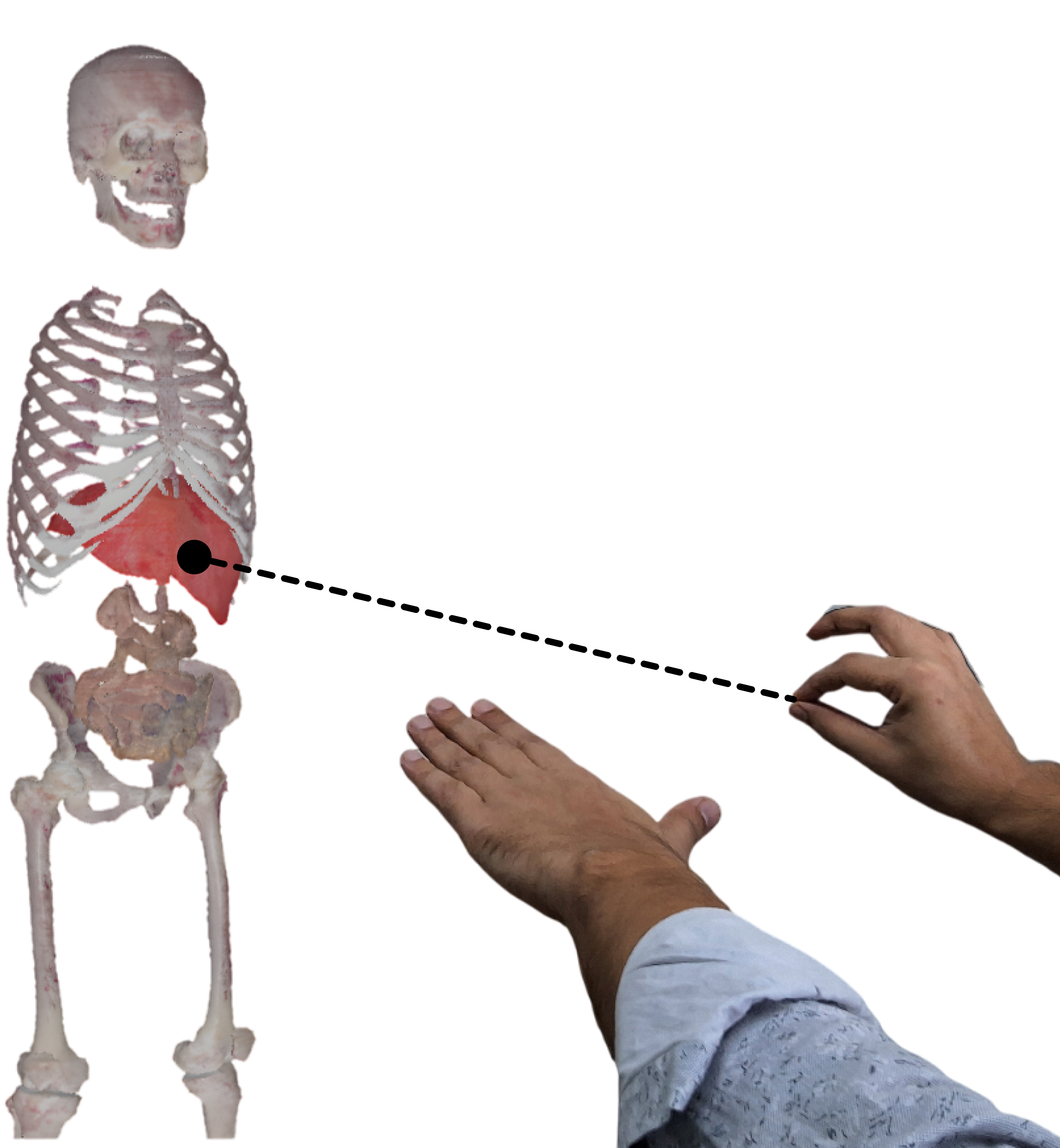}\\
    (a) \emph{Clipping plane} & (b) \emph{Selective rendering} & (c) \emph{Organ information} & (d) \emph{Bioscope}
    \end{tabular}
    \caption{Holoview interaction features. (a) Hand-controlled clipping plane for cross-sectional volume visualization. (b) Selective rendering of L$_2$ organs via gesture-based selection. (c) Organ information retrieval using a pointing fist gesture. (d) Bioscope mode activated by a two-hand gesture for immersive organ inspection.}
    \label{fig:features}
\end{figure*}

\subsubsection{Clipping Plane}
Cross-sectional views are essential in medical diagnostics for analyzing internal organ structures, similar to those in CT or MRI scans. Holoview enables this through an interactive clipping plane that allows users to slice through volumetric anatomy in real time.

The clipping plane is activated through the UI and attaches to the user's right hand when held straight with a visible gap between the index finger and thumb as shown in Figure~\ref{fig:features}(a). The orientation of the plane continuously follows the position of the hand and the normal vector, allowing users to move, rotate, and explore the internal structures precisely. Further the orientation is in sync with the user’s gaze, automatically ensuring that the side facing the user is clipped. When the gesture is released, the plane remains fixed in its last position, locking the view for further examination.


\subsubsection{Selective Rendering}
\label{ssec:selective_rendering}

In clinical diagnostics and anatomy education, the ability to isolate specific organs or tissues for focused analysis is essential whether interpreting complex scans or understanding spatial relationships. Holoview addresses this through selective rendering, enabling users to interactively visualize one or more organs within a complex anatomical volume.

Selection begins at the L$_1$ category level using UI buttons. Once a system is chosen, its L$_2$ sub-organs are revealed and color-coded for clarity (see Figure~\ref{fig:features}(b)). Users can select or deselect individual L$_2$ organs by performing a pinch gesture with their right hand—bringing the index finger and thumb together while pointing at the target organ. This same gesture toggles the selection state. Selected meshes are visually highlighted, and users can also select or deselect all sub-organs in a system via the UI. Holoview supports simultaneous multi-system selection for flexible composition.

\subsubsection{Label Information}
Holoview enables users to identify and learn about anatomical structures through a simple one-handed gesture. To activate this feature, the user forms a fist with their right hand, pointing it at the target organ within the camera’s view  as shown in Figure~\ref{fig:features}(c). Upon activation, the organ name appears. Holding the gesture slightly longer reveals additional functional details. This gesture can be performed at any time and works alongside other interactions, offering quick, contextual access to anatomical information.


\subsubsection{Navigation}
Holoview implements a driving-style navigation model that allows users to explore anatomical structures while remaining physically stationary. This is achieved by translating the body model in the opposite direction of the user's gesture, creating the illusion of movement relative to the anatomy. For example, a leftward gesture moves the model to the right, making the user feel as though they have moved left. This approach enables immersive internal navigation within limited physical spaces, enhancing comprehension of spatial anatomical relationships.

\begin{figure}[t]
\begin{tabular}{ll} 
\includegraphics[width=0.45\columnwidth,keepaspectratio]{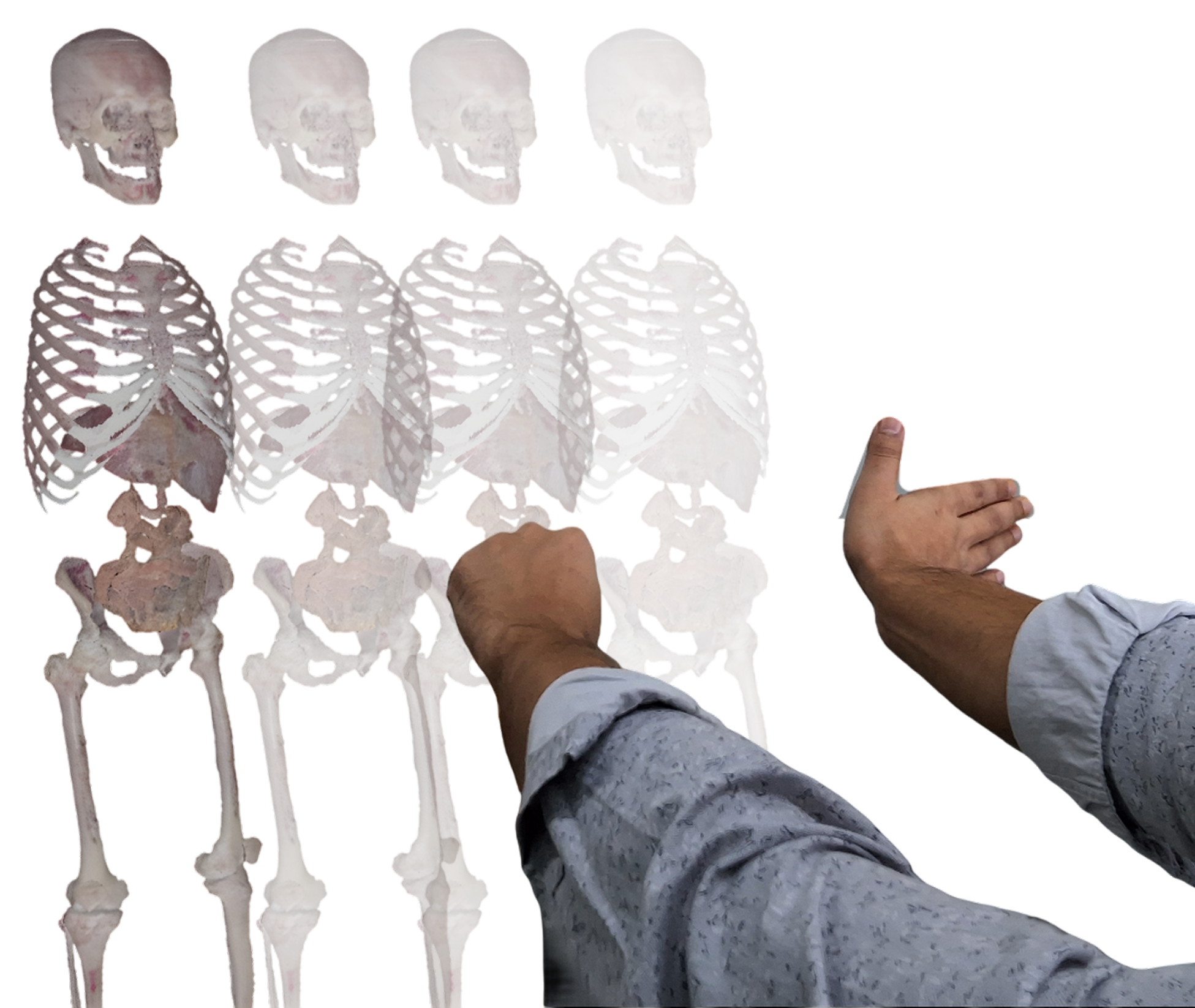} &
\includegraphics[width=0.45\columnwidth,keepaspectratio]{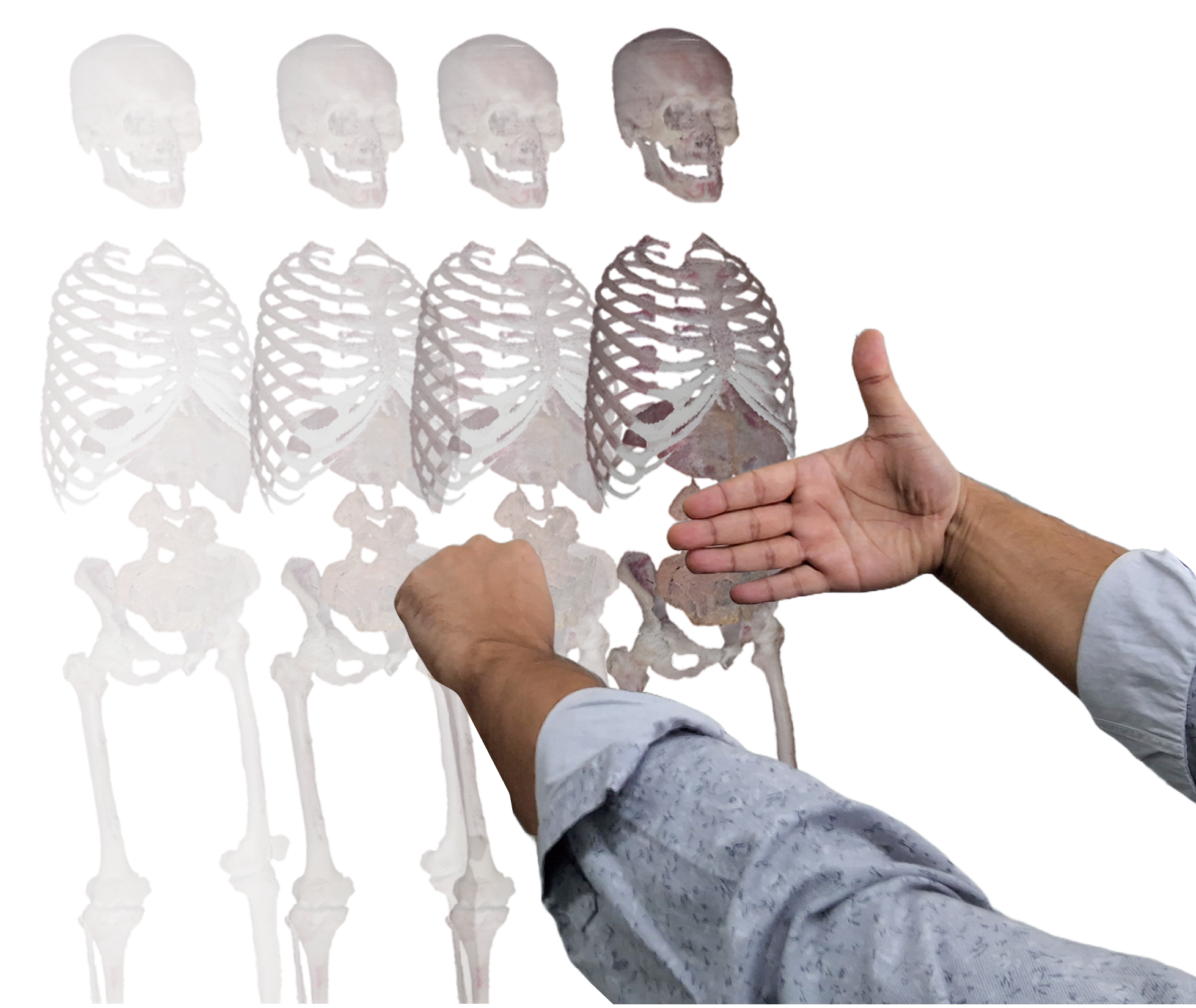} \\
\includegraphics[width=0.45\columnwidth,keepaspectratio]{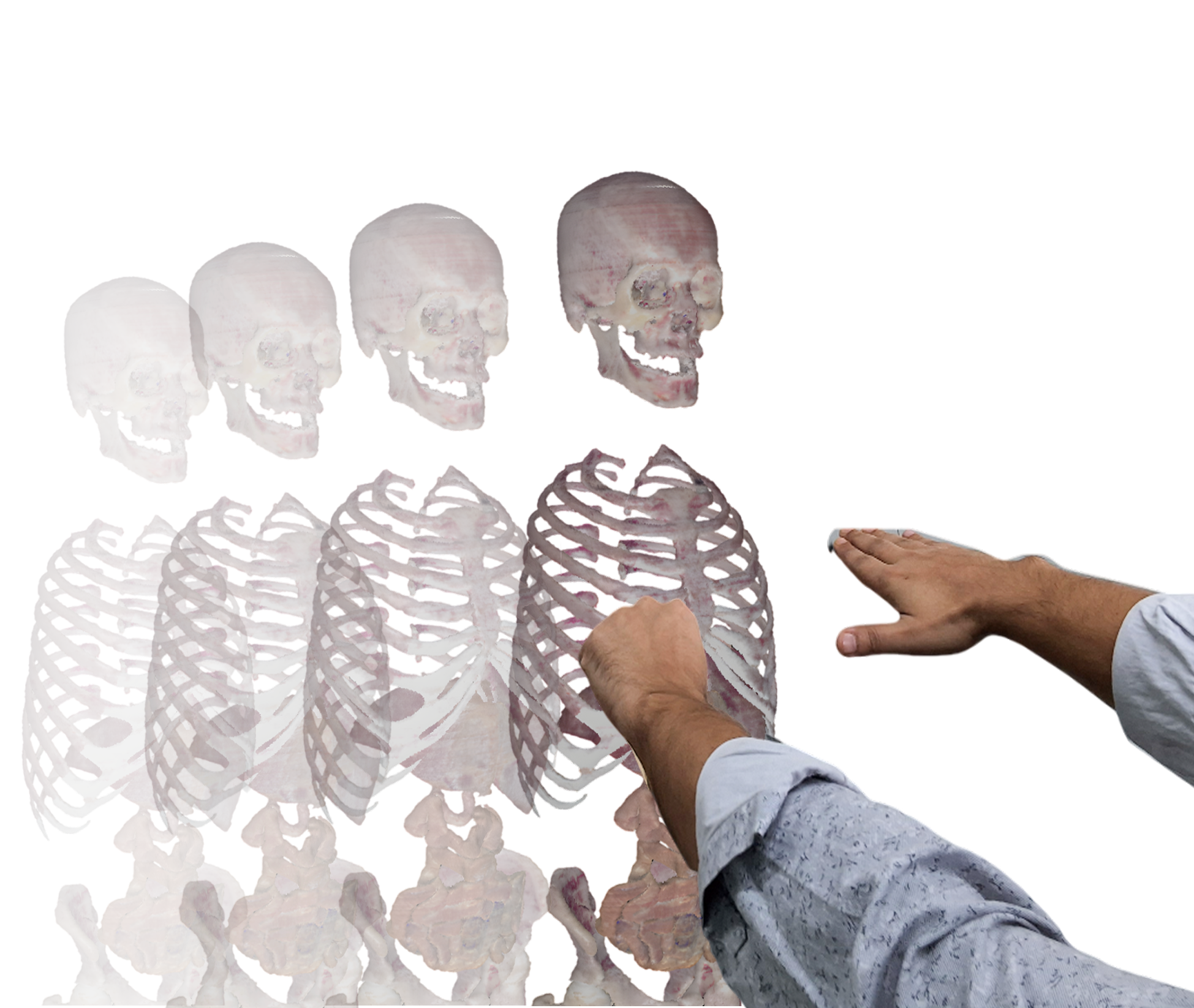} & 
\includegraphics[width=0.45\columnwidth,keepaspectratio]{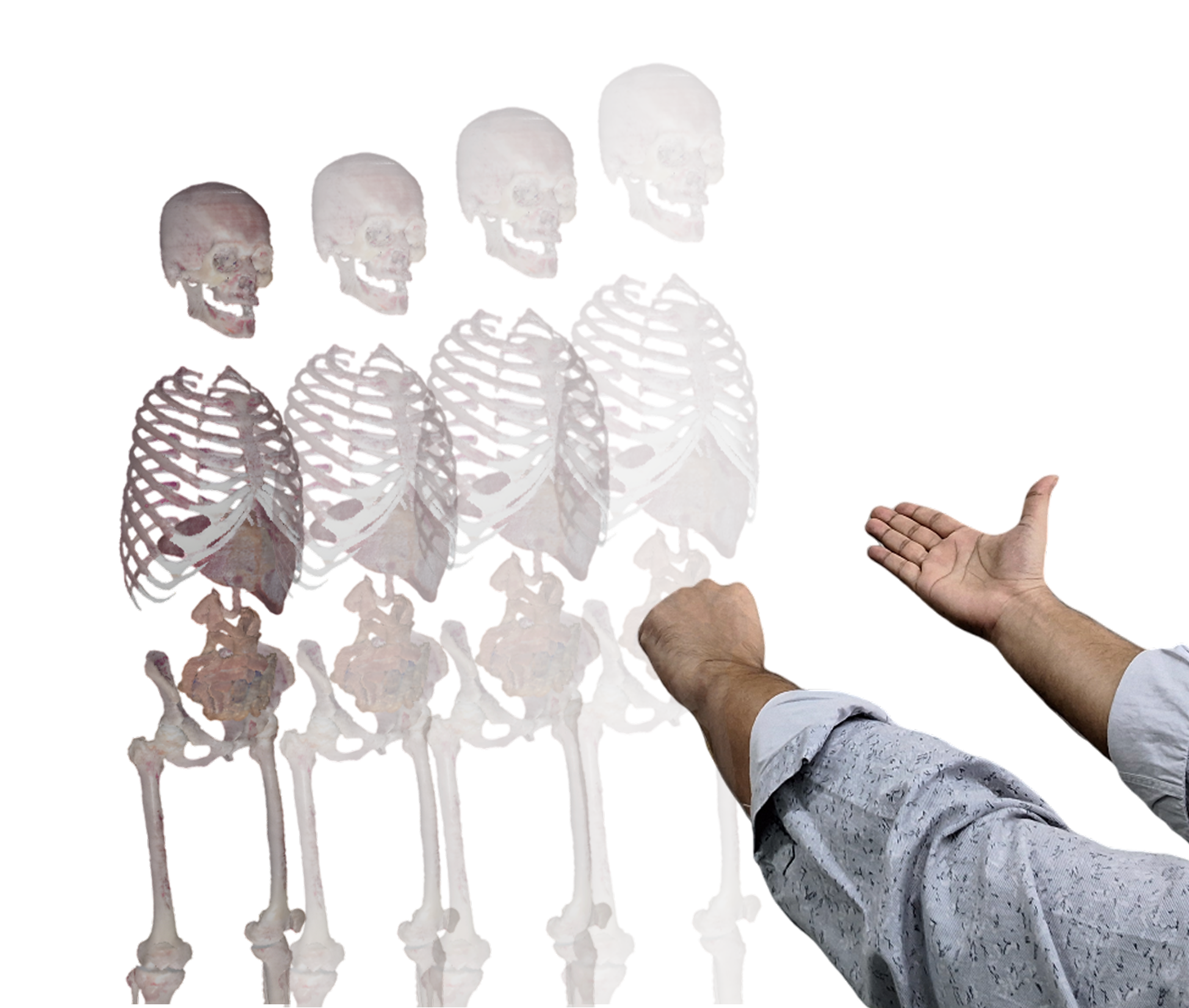} \\
\includegraphics[width=0.45\columnwidth,keepaspectratio]{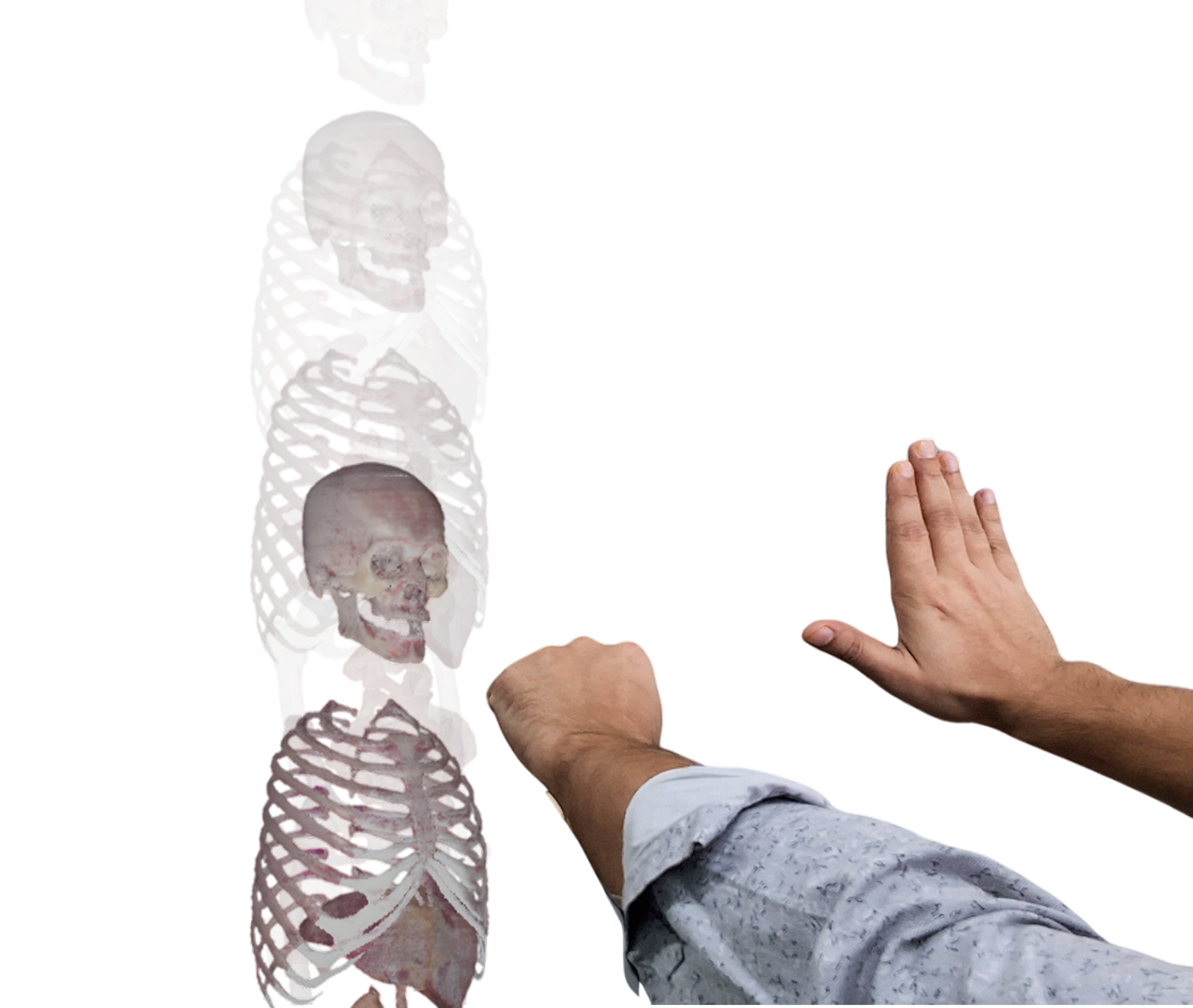} &
\includegraphics[width=0.45\columnwidth,keepaspectratio]{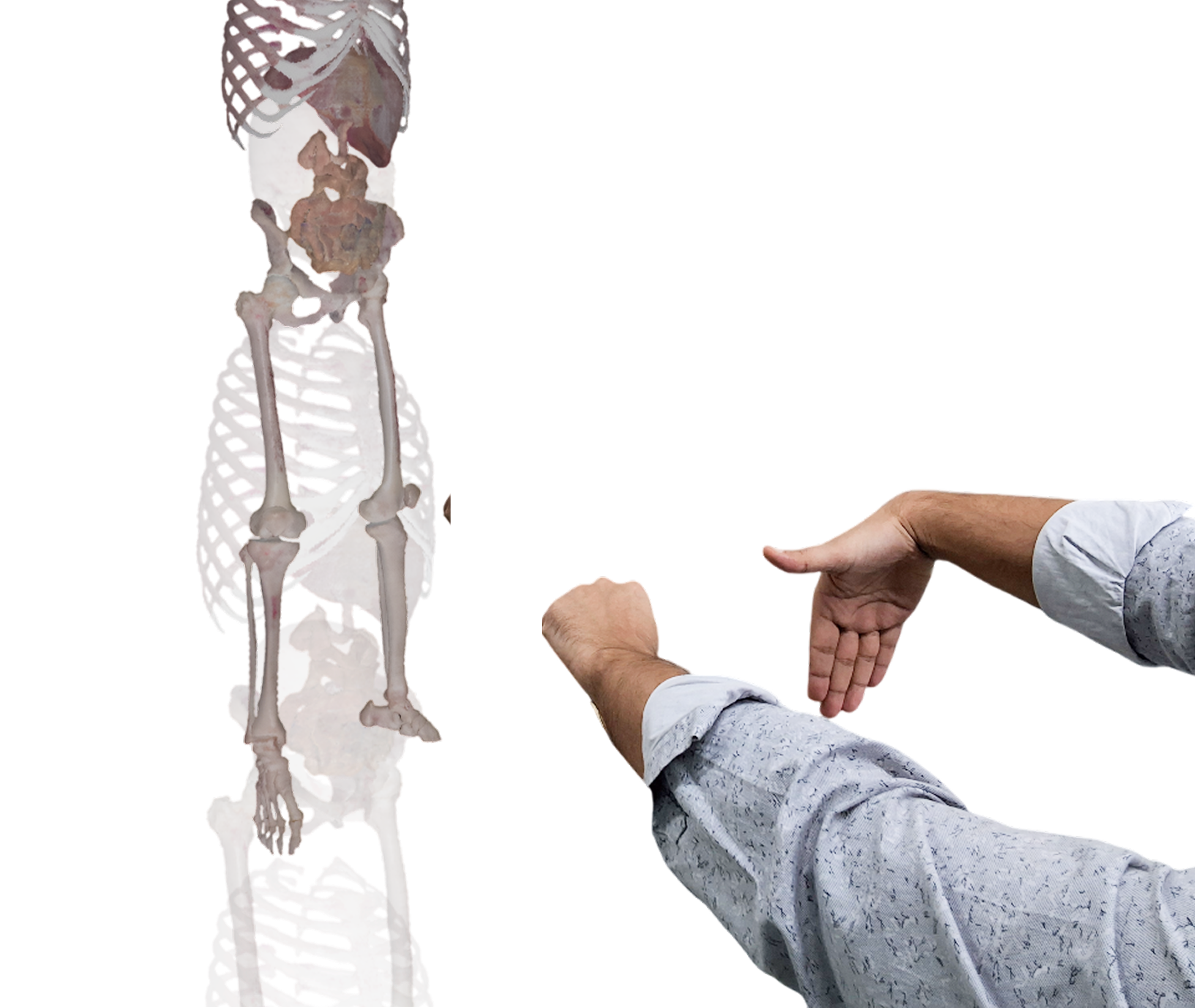}         
\end{tabular}
\caption{Navigation gestures in Holoview, where the right hand controls camera movement. Top row (left to right): move right, move left. Middle row (left to right): move forward, move backward. Bottom row (left to right): move up, move down.}
\label{fig:navGesture}
\end{figure}

Navigation is activated using a two-handed gesture: the user makes a fist with their left hand (in view of the camera) to enter navigation mode, while the flat right hand indicates the movement direction (see Figure~\ref{fig:navGesture}). Six directions are supported—forward, backward, left, right, up, and down—mapped to specific orientations of the right hand. To ensure smooth interaction, Holoview modulates navigation speed based on the user's proximity to the model: maintaining a constant speed when distant and reducing it exponentially when close. During navigation, anatomical labels appear contextually near relevant structures, reinforcing learning through spatial cues and gesture-driven interaction.

\subsubsection{Bioscope}
While Holoview’s Navigation mode enables full-body exploration, the Bioscope mode offers a focused view for detailed inspection of a single organ’s internal anatomy. This mode is designed to enhance understanding of complex structures within an organ.

To activate Bioscope, the user performs a two-hand gesture: the left hand is held flat with the palm facing downward, while the right hand executes a pinch gesture aimed at the target organ as shown in Figure~\ref{fig:features}(d). Upon recognition, the selected organ smoothly animates into position directly in front of the user, scaling to twice its original size for enhanced visibility. In this mode, users can access all major functionalities, including the clipping plane, spatial navigation, and manipulation tools such as translation and rotation. These interactions remain consistent with other modes, supporting a seamless user experience.



\section{User Study and Evaluation}

We conducted an Institutional Review Board-approved user study to evaluate the efficacy of our AR-based system designed to enhance medical training, with a specific focus on organ-based inspection and identification tasks. By comparing the system to conventional educational tools, including medical textbooks and on-screen software, the study aimed to delineate its strengths and areas for improvement. The findings offer valuable insights into the integration of AR technologies within classroom learning environments to elevate the quality of medical education.

\paragraph{Hypothesis}
We hypothesize that our system significantly enhances the quality of medical training and learning experiences by providing an intuitive, accessible platform that reduces cognitive load compared to traditional educational methodologies.

\subsection{Participants}
Fifteen participants were recruited using snowball sampling, including 11 undergraduate, 2 postgraduate, and 2 doctoral students (11 males, 4 females; mean age: 24 years) from technical backgrounds such as biomedical engineering, computer science, and related disciplines. All participants had completed school-level education with biology and human anatomy as part of the curriculum. This ensured a foundational familiarity with anatomical structures, enabling them to meaningfully participate in spatial and organ-identification tasks. Two participants had prior exposure to AR technologies (0–1 years), while the rest had minimal to no experience.


\subsection{Procedure}
In collaboration with medical professionals, we developed three comprehensive questionnaire sets, targeting liver, heart, and skull. Each set included pre-task and post-task assessments to evaluate demographics, spatial intelligence, and organ-specific knowledge. Participants provided informed consent prior to participation. The study commenced with an orientation and pre-training session to familiarize users with the system's features. Ethical considerations, including confidentiality and privacy, were rigorously upheld.

Following orientation, each participant completed a pre-task questionnaire, which included two organ-specific questions to assess their baseline knowledge of the organs based on traditional learning methods. The questionnaire also gathered demographic information, including age, educational background, and prior AR/VR experience. Once the pre-task questionnaire was completed, participants were asked to perform the following main tasks using our system:

\begin{itemize} 
    \item Explore, identify, and locate the L$_2$ segments of the specified organs,
    \item Inspect and explore the neighboring organs and surrounding anatomy, and
    \item Examine the specified organ, gather its name and information, and analyze both its external and internal structures to deepen their understanding. 
\end{itemize}

Upon task completion, participants completed a post-task questionnaire assessing task performance, system usability, and cognitive workload. The questionnaire included:

\begin{itemize} 
\item Task-specific assessments (pre- and post-task comparisons),
\item System feature evaluation (ease of interaction with holographic models),
\item System Usability Scale (SUS) \cite{sus} for user-friendliness evaluation, and
\item Simulation Task Load Index (SIM-TLX) \cite{Harris2020} for cognitive workload assessment.
\end{itemize}

\subsection{Analysis}
We analyzed the performance of participants in completing the tasks, their learning progress, and feedback on the system's features and usability.

\subsubsection{Learning Outcomes}
Learning outcomes were assessed by comparing pre-task and post-task questionnaire scores. The mean pre-test score was 2.60 (SD = 1.96), increasing to 6.13 (SD = 2.07) post-test, reflecting a 135\% improvement in task-specific knowledge. A paired t-test confirmed the statistical significance of this improvement ($t(14) = 5.89$, $p < 0.001$). Figure \ref{fig:knowledge_improvement} illustrates individual improvements.

Participants reported increased confidence in identifying and understanding the internal and external structures of organs, which is critical in medical training. The immersive AR experience likely facilitated enhanced visualization and memorization of complex anatomical relationships.

\begin{figure}[!h] 
    \centering 
    \begin{tikzpicture}    
    \begin{axis}[
    cycle list name=color list, 
    ybar=0pt,
    bar width=5pt,
    title={\small \textbf{Task specific knowledge}},    
    height=4cm, 
    width=\linewidth,
    area style,
    axis lines=left, 
    axis line style={-{Latex[round]}},
    ymin=0,ymax=11,
    xmin=0.5,xmax=16,
	xtick={1,2,3,4,5,6,7,8,9,10,11,12,13,14,15},
    xticklabels={1,2,3,4,5,6,7,8,9,10,11,12,13,14,15},
    ylabel={\small \emph{Score}},
    xlabel={\small \emph{Participant number}},
    legend style={at={(0.5,1.0)},anchor=north,legend columns=-1, fill=gray!30!white, draw=none},
    ymajorgrids=true,        
	ytick distance=2,
    every axis plot/.append style={fill,fill opacity=0.75},
    x tick label style={rotate=0, font=\small},  
    y tick label style={font=\small}, 
    ]
    \pgfplotsset{cycle list shift=+2}
    \addplot+[ybar] table[row sep=\\, x index=0, y index=1] { 
    1 2 \\ 2 3 \\ 3 3 \\ 4 4 \\ 5 6 \\ 6 1 \\ 7 3 \\ 8 1 \\ 9 0 \\ 10 0 \\ 
    11 4 \\ 12 2 \\ 13 5 \\ 14 0 \\ 15 5 \\ };
	\addplot+[ybar] table[row sep=\\, x index=0, y index=1] { 
    1 2 \\ 2 5 \\ 3 4 \\ 4 8 \\ 5 7 \\ 6 8 \\ 7 6 \\ 8 7 \\ 9 8 \\ 10 3 \\ 
    11 8 \\ 12 6 \\ 13 6 \\ 14 5 \\ 15 9 \\ };

    \legend{\small Pre-test, \small Post-test}					
    \end{axis}
	\end{tikzpicture}
    \caption{Pre- and post-test score comparison per participant.} 
    \label{fig:knowledge_improvement} 
\end{figure}

\subsubsection{System Usability Evaluation}
The system achieved a mean System Usability Scale (SUS) score of 68.33 (SD = 16.29), indicating moderate usability. Scores ranged from 42.5 to 90, with a median of 72.5 and an interquartile range of 56.25 to 78.75. Figure \ref{fig:sus_score} presents the distribution of SUS scores.

Participants with prior AR experience adapted quickly to the system, while those unfamiliar with AR encountered an initial learning curve, particularly with gesture-based interactions. Approximately 80\% of participants had no prior exposure to AR, and this group showed initial uncertainty about gesture-function mappings, likely contributing to the moderate SUS scores. In contrast, experienced users reported no difficulty with the gesture interface and achieved higher SUS scores, indicating that prior familiarity with AR significantly reduces usability barriers. This group adapted more rapidly and interacted with the system more fluidly.

Across all participants, usability improved notably over time. Repeated interaction led to greater confidence and efficiency, with several users remarking that tasks became ``easier to use'' as they progressed. This trend underscores the role of practice in overcoming the initial learning curve. These findings suggest that while gesture-based interaction may initially hinder usability for AR novices, familiarity develops quickly. Future system iterations could benefit from onboarding tutorials or adaptive guidance to ease this transition for new users.

\begin{figure}[h!] 
    \centering 
\begin{tikzpicture}
    \begin{axis}[
        cycle list name=color list, 
        ybar interval,
    	title={\small \textbf{Distribution of SUS scores}},    
	    height=4cm, 
    	width=\linewidth,        
        ymin=0,
        ymax=5, 
        xmin=0,
        xmax=100,
    	axis lines=left, 
	    axis line style={-{Latex[round]}},        
	    xmajorgrids=false,
	    ymajorgrids=true,        
		ytick distance=1,
        xticklabel=\pgfmathprintnumber\tick--\pgfmathprintnumber\nexttick,
        xlabel={\small \emph{Score}},
        ylabel={\small \emph{Frequency}},
	    x tick label style={rotate=90, font=\small},  
    	y tick label style={font=\small}, 
     ]
        \pgfplotsset{cycle list shift=+2}
        \addplot+[hist={data min=0,data max=100,bins=20}] table[row sep=\\, y index=0] {
        	data\\
            70 \\ 60 \\ 90 \\ 60 \\ 72.5 \\ 52.5 \\ 80 \\ 72.5 \\ 45 \\ 42.5 \\
            50 \\ 90 \\ 77.5 \\ 90 \\ 72.5 \\
        };
    \end{axis}
\end{tikzpicture}
    \caption{System Usability Scale (SUS) scores for the system.} 
    \label{fig:sus_score} 
\end{figure}

\subsubsection{Cognitive Load and Task Demand}
The mean Task Load Index (TLX) score for the system was 36.4, reflecting a low cognitive load and indicating that tasks were manageable without excessive mental strain. Participants with no prior AR experience initially found gesture-based interactions challenging, but perceived effort decreased with familiarity. Figure \ref{fig:tlx_scores} displays boxplots for each TLX dimension, highlighting variations in cognitive demand across mental fatigue, physical fatigue, and other factors.

Participants with prior AR experience reported lower cognitive load, aligning with their familiarity with similar interaction paradigms. Across the cohort, no participants reported feelings of being overwhelmed or excessively frustrated. Many remarked that the tasks were engaging and that the learning curve, while present, did not lead to significant mental fatigue. The results shown in \ref{fig:tlx_scores} indicate that the system effectively balances cognitive demand, making it accessible to both new and experienced users with minimal adaptation time.

\begin{figure}[h!] 
    \centering 
    \begin{tikzpicture}
    \begin{axis}[
        cycle list name=color list, 
        boxplot/draw direction=y,
        height=5cm,
        width=\linewidth,
        title={\small \textbf{Cognitive load during the task}},    
        ymin=0, ymax=10,
        axis lines=left,
        axis line style={-{Latex[round]}},
        ylabel={\small \emph{Score}},          
        boxplot={
  	      	box extend=0.5,
        	every median/.style=ultra thick
        },
        xmin=0.5, xmax=10.5,
        xtick={1,2,3,4,5,6,7,8,9,10},
        xticklabels={MF,PF,HU,IR,TC,ST,DE,US,TD,IM},
        every axis plot/.append style={fill,fill opacity=0.25},
	    x tick label style={rotate=0, font=\small},  
    	y tick label style={font=\small},        
        ]
        \pgfplotsset{cycle list shift=4}
        \addplot+ [boxplot, mark=*]  table[row sep=\\,y index=0] {
        data\\6\\2\\3\\9\\4\\1\\2\\3\\9\\2\\6\\1\\2\\2\\4\\};
        \addplot+ [boxplot, mark=*]  table[row sep=\\,y index=0] {
        data\\3\\2\\3\\7\\8\\1\\3\\3\\6\\7\\4\\1\\3\\5\\8\\};
        \addplot+ [boxplot, mark=*]  table[row sep=\\,y index=0] {
        data\\2\\2\\1\\7\\2\\1\\1\\3\\6\\6\\4\\1\\1\\2\\3\\};
        \addplot+ [boxplot, mark=*]  table[row sep=\\,y index=0] {
        data\\2\\1\\2\\3\\2\\2\\2\\2\\7\\3\\3\\1\\2\\2\\2\\};
        \addplot+ [boxplot, mark=*]  table[row sep=\\,y index=0] {
        data\\4\\4\\1\\7\\3\\2\\1\\1\\9\\8\\6\\2\\4\\5\\3\\};
        \addplot+ [boxplot, mark=*]  table[row sep=\\,y index=0] {
        data\\3\\3\\2\\3\\3\\1\\2\\1\\7\\4\\3\\1\\2\\1\\1\\};
        \addplot+ [boxplot, mark=*]  table[row sep=\\,y index=0] {
        data\\7\\2\\2\\3\\1\\1\\4\\3\\9\\4\\3\\1\\2\\1\\1\\};
        \addplot+ [boxplot, mark=*]  table[row sep=\\,y index=0] {
        data\\6\\1\\2\\2\\2\\1\\2\\1\\8\\2\\4\\1\\1\\1\\2\\};
        \addplot+ [boxplot, mark=*]  table[row sep=\\,y index=0] {
        data\\5\\7\\1\\4\\4\\1\\1\\4\\9\\7\\6\\2\\3\\5\\5\\};
        \addplot+ [boxplot, mark=*]  table[row sep=\\,y index=0] {
        data\\7\\2\\9\\7\\8\\1\\7\\5\\8\\8\\9\\9\\9\\10\\9\\};                                                                                              
    \end{axis}
    \end{tikzpicture}    
    \caption{SIM-TLX scores reflecting cognitive load during the task. MF: Mental Fatigue, PF: Physical Fatigue, HU: Hurried, IR: Insecure/Irritated, TC: Task Complexity, ST: Stress, DE: Distracting Environment, US: Uncomfortable Sensory, TD: Task Difficulty, IM: Immersion.} 
    \label{fig:tlx_scores} 
\end{figure}

\subsubsection{Feature Feedback}
Participants provided detailed feedback on system features, highlighting both strengths and areas for enhancement.\\
\textbf{Clipping Plane}: The clipping plane was well-received for its intuitive control, ease of use, and ability to display clear and informative cross-sections of anatomical structures. Most participants found the controls intuitive for exploring specific areas of the body, though a few noted some limitations in flexibility.\\
\textbf{Organ Selection}: Many participants appreciated the feature of selecting specific organs from groups and isolating them for closer study. Some proposed further dividing organs into smaller sections for more detailed examination.\\
\textbf{Bioscope Mode}: The bioscope mode was widely praised for providing detailed views of selected organs. Participants found adjusting the bioscope level straightforward, enabling them to explore both internal and external structures with ease. The scaled-up view of organs was noted as particularly effective for understanding anatomy in greater detail.\\
\textbf{Gesture Controls and Navigation}: Feedback on gesture controls was mixed. While many participants found the gestures intuitive and effective for navigation, others encountered issues with responsiveness, especially for complex gestures activating the bioscope mode. Navigation in six directions was generally smooth, but a minority of users, especially those new to AR systems, found it initially confusing. Prolonged use of hand gestures was comfortable for most participants, though a few mentioned experiencing mild fatigue during extended sessions.

Commonly suggested improvements included refining gesture recognition, enhancing the interface's visual clarity (e.g., increasing text readability), and adding more vivid or detailed visualizations. Participants also proposed dividing organ structures into finer components for deeper exploration and incorporating features like automatic centering for ease of navigation.

Overall, participants rated the system positively, with many noting that the features, particularly clipping and organ selection, provided an engaging and effective way to explore anatomy. These findings emphasize Holoview's potential for supporting educational applications, while also identifying areas where usability could be improved for new and experienced users alike.

\section{Conclusions}
We developed Holoview, an immersive AR system designed to facilitate the study of human anatomy for medical students. The application utilises foveated rendering, enabled by the eye-tracking capabilities inherent in modern AR devices, to optimize rendering efficiency and enhance the user experience. To eliminate the need for controllers, we implemented intuitive one-hand and two-hand based gesture interactions. A single-hand gesture allows users to access detailed organ information, while two-hand gestures enable navigation within constrained physical environments.

The system also includes a bioscope mode, which features a simple slider to adjust the transparency of anatomical boundaries, providing users with an interactive way to explore internal structures. To assess its effectiveness, we conducted a user study that focused on usability, the intuitiveness of gesture-based interactions, and the quality of the 3D visualization of anatomical data. The results highlight the system's potential as a valuable tool for enhancing anatomy education through immersive AR technology.


\begin{acks}
This research was supported by the Science and Engineering Research Board (SERB) of the Department of Science and Technology (DST) of India (Grant No. CRG/2020/005792). The Visible Korean Human dataset for our research was provided by  the Korea Institute of Science and Technology Information (KISTI), South Korea.
\end{acks}

\printbibliography

\end{document}